\documentclass[journal,12pt,onecolumn,draftclsnofoot]{IEEEtran}
\ifCLASSINFOpdf

\else

\fi
\usepackage[colorinlistoftodos]{todonotes}
\usepackage{amssymb}
\usepackage{amsmath,epsfig,amssymb,verbatim}
\usepackage{cite}
\usepackage[caption=false]{subfig}
\usepackage{comment}
\usepackage{psfrag}
\usepackage{stfloats}
\usepackage{footnote}
\usepackage{multirow}
\usepackage{multicol}
\usepackage[T1]{fontenc}
\def\BibTeX{{\rm B\kern-.05em{\sc i\kern-.025em b}\kern-.08em
    T\kern-.1667em\lower.7ex\hbox{E}\kern-.125emX}}

\newtheorem{theorem}{Theorem}
\newtheorem{corollary}{Corollary}
\newtheorem{lemma}{Lemma}
\newtheorem{proposition}{Proposition}
\newtheorem{remark}{Remark}

\def\IEEEproofname{Proof}

\def\IEEEQEDclosed{\mbox{\rule[0pt]{1.3ex}{1.3ex}}} 

\def\IEEEQED{\IEEEQEDclosed} 

\def\IEEEproof{\@ifnextchar[{\@IEEEproof}{\@IEEEproof[\IEEEproofname]}}
\def\@IEEEproof[#1]{\@IEEEQEDshowtrue\par\noindent\hspace{\IEEEproofindentspace}{\itshape #1: }}
\def\endIEEEproof{\if@IEEEQEDshow\hspace*{\fill}\nobreakspace\IEEEQED\fi\par}

\begin{document}
\bstctlcite{IEEEexample:BSTcontrol}

\title{Coverage Analysis for 3D Terahertz Communication Systems}

\author{Akram~Shafie, Nan~Yang, Salman~Durrani, Xiangyun Zhou, Chong~Han, and~Markku Juntti 
\thanks{ This article was presented in part at the IEEE ICC 2020 Workshop on Terahertz Communications \cite{akramICCWS2020}.}
\thanks{The work of Nan Yang and Xiangyun Zhou was supported by the Australian Research Council Discovery Project under Grant DP180104062. The work of Chong Han was supported by the National Key Research and Development Program of China under Project 2020YFB1805700. The work of Markku Juntti was supported in part by the Academy of Finland through 6Genesis Flagship under Grant 318927. }
\thanks{A. Shafie, N. Yang, S. Durrani, and X. Zhou are with the School of Engineering, The Australian National University, Canberra, ACT 2600, Australia (Emails: \{akram.shafie, nan.yang, salman.durrani, xiangyun.zhou\}@anu.edu.au).}
\thanks{C. Han is with the UM-SJTU Joint Institute, Shanghai Jiao Tong University, Shanghai 200240, China (Email: chong.han@sjtu.edu.cn).}
\thanks{M. Juntti is with the Centre for Wireless Communications, University of Oulu, Oulu 90014, Finland (Email: markku.juntti@oulu.fi). }
}

\markboth{}
%
{Shell \MakeLowercase{\emph{et al.}}: Bare Demo of IEEEtran.cls for IEEE Journals}

\maketitle

\enlargethispage{0.3 cm}

\vspace{-16mm}
\begin{abstract}
\vspace{-3mm}
We conduct novel coverage probability analysis of downlink transmission in a three-dimensional (3D)  terahertz (THz) communication (THzCom) system. In this system, we address the unique propagation properties in THz band, e.g., absorption loss, super-narrow directional beams, and high vulnerability towards blockage,  which are fundamentally different from those at lower frequencies. Different from existing studies, we characterize the performance while considering the effect of 3D directional antennas at both access points (APs) and user equipments (UEs), and the joint impact of the blockage caused by the user itself, moving humans, and wall blockers in a 3D environment. Under such consideration, we develop a tractable analytical framework to derive a new expression for the coverage probability by examining the regions where dominant interferers (i.e., those can cause outage by themselves) can exist, and the average number of interferers existing in these regions. Aided by numerical results, we validate our analysis and reveal that ignoring the impact of the vertical heights of THz devices in the analysis leads to a substantial underestimation of the coverage probability. We also show that it is more worthwhile to increase the antenna directivity at the APs than at the UEs, to produce a more reliable THzCom system.

\end{abstract}
\vspace{-5mm}

\IEEEpeerreviewmaketitle

\section{Introduction}

\enlargethispage{0.3 cm}

Terahertz (THz) communication (THzCom)  has been  envisaged as a highly promising  paradigm to alleviate the spectrum scarcity and break the capacity limitation of contemporary wireless networks \cite{20206GMag}. In particular, the ultra-wide THz band ranging from 0.1 to 10 THz provides enormous potential to realize sixth-generation (6G) wireless network applications that demand high quality of service requirements and multi-terabits per second data transmission. These applications, such as ultra-fast wireless local area networks and wireless virtual/augmented reality, are beyond the reach of sub-6~GHz and millimeter wave (mmWave) communication, which undoubtedly creates the need of THzCom \cite{2019_OJCS_Survey_SalmanRef1}. Built on the major progress in THz hardware design, e.g., new graphene-based THz transceivers and ultra-broadband antennas that operate at THz frequencies \cite{2017_ChongTVT_Graphene}, and the THzCom standardization efforts  over the past decade \cite{2020_WCM_THzMag_Standardization}, it is anticipated that indoor THzCom systems will be brought to reality in the near future.

Despite the promise, the design of ready-to-use THzCom systems brings new and pressing challenges that have never been seen at lower frequencies  \cite{2014_ChongMag}.
For example, the THz band suffers from very high spreading loss and highly frequency-selective molecular absorption loss which profoundly decreases the THz transmission distance \cite{2011_Jornet_TWC}.
Moreover, the high reflection and scattering losses  significantly attenuate the non-line-of-sight (NLoS) rays \cite{HBM3}. 
Furthermore, the  extremely short wavelength of THz signals makes the THz signal propagation to be highly vulnerable to blockages \cite{2018MagCombatDist}. Specifically, objects with small dimensions such as the user itself, moving humans, and inherent indoor constructions (e.g., walls and furniture), can act as impenetrable blockers. Thus, the propagation environment at the THz band is unique, which motivates the design and development of new communication paradigms and novel signal processing tools to tackle these challenges.

In THzCom systems, narrow beams that have high directional gains are to be utilized at  transceivers to compensate for the severe path loss.
Fortunately, it is possible to integrate multiple antennas into THz transceivers to form narrow beams due to the extremely short wavelength at the THz band \cite{2020_Kurner_2Vehicle,2017_ChongTVT_Graphene,2020_TC_HangNan}.
Although the use of very narrow beams may eventually lead to the noise-limited regime of wireless communication, the state-of-the-art THz antennas produce beams that have reasonably large beamwidths to cause interference in a multi-user network \cite{2020_Kurner_2Vehicle}.
Moreover, the increase in network densification, the use of advanced networking mechanisms, and device-to-device communication are likely to increase the interference in THzCom systems
\cite{2018MagLastMeterIndoor,akramICC2020,2020_WCL_RISforTHz}.
Furthermore, narrow beams along with blockages introduce fundamentally different interference patterns  at the THz band  from those observed at lower frequencies. Specifically, at the THz band, a lesser number of interferers contribute to the aggregated interference, while the impact of each interferer that contributes to the aggregated interference is significant.
\textit{Therefore, assessing the reliability of THzCom systems in the presence of interference is of great significance for developing ready-to-use THzCom systems.}

Conventionally, the coverage probability is used as a key reliability performance metric. In sub-6 GHz and mmWave communication systems, the coverage probability in the presence of interference has been widely investigated \cite{LaplaceIntRef2,StochGeo2011,Salman2014Stoch}.
Despite this, the studies on the coverage probability at the THz band that incorporates all the unique characteristics of THzCom systems are very limited.  
Using tools from stochastic geometry and considering interference limited regime, the coverage probability for a THzCom system was determined in \cite{2017M12} by approximating the interference distribution using a log-logistic distribution.
The interference and coverage probability of a co-existing sub-6 GHz and dense THz  wireless network was investigated in \cite{2020_WCL_Multi_RATMCforTHz}, using the conventional Laplace transform-based analysis.
In addition, the interference distribution was approximated using a normal distribution in \cite{2015_VTC_THzIntNormalDist} for tractable coverage probability analysis of dense THzCom systems.
Moreover, coverage probability of downlink indoor THzCom systems was evaluated in \cite{CovTHz2019ICCC} while considering the impact of both line-of-sight (LoS) and NLoS rays. The studies in \cite{2017M12,2020_WCL_Multi_RATMCforTHz,2015_VTC_THzIntNormalDist,CovTHz2019ICCC} considered the impact of directional antennas, but did not examine the impact of blockages which can greatly affect the reliability performance of THzCom systems.

The joint effect of human blockages and directional antennas on the THzCom  system performance was studied in \cite{2017M5,2018NN3}.
Specifically, the mean and variance of interference and signal-to-interference-plus-noise ratio (SINR) were determined using the Taylor approximation in \cite{2017M5}. Analytical approximations of the interference and signal-to-interference ratio were presented in \cite{2018NN3}. However, the studies in \cite{2017M5,2018NN3}  only derived the first few moments of the metric of interest. In other studies, the coverage probability of THzCom systems was derived in \cite{2017O1,2017N3}  while considering the joint impact of blockages and directional antennas. However, the accuracy of their results deteriorates significantly, especially for long transmission distance, since they used the average interference instead of the instantaneous interference when evaluating the coverage probability. In addition, the prior work in \cite{2017M12,2020_WCL_Multi_RATMCforTHz,2015_VTC_THzIntNormalDist,CovTHz2019ICCC,2017M5,2018NN3,2017N3,2017O1} focused on a two-dimensional (2D) environment only.

In sub-6 GHz and mmWave systems, it may be reasonable to ignore the impact of the vertical height of communication entities  due to the large transmission distance. However, the vertical dimension may greatly impact the reliability performance of THzCom systems,  since transmission distances at the THz band are limited to the order of few meters. \textit{Thus, it is essential to consider the vertical height of communication entities and investigate its impact on the coverage probability in THzCom systems.} Recently, the coverage probability  for a THzCom system was determined in a three-dimensional (3D) environment in \cite{2020_ChongCoverage}, but it did not address the impact of 3D directional antennas at both transmitter and receiver sides.
In addition, it has to be noted that the prior studies in \cite{2017M12,2020_WCL_Multi_RATMCforTHz,2015_VTC_THzIntNormalDist,CovTHz2019ICCC,2017M5,2018NN3,2017N3,2017O1} did not consider the impact of wall blockers, despite that the primary applications of THzCom systems are found in indoor environment where walls exist.

\enlargethispage{0.25 cm}

In this paper\footnote{The analysis in our preliminary work in [1] was performed for an open office environment where only human blockers exist. In [1], the human blockers were modelled using a simplified circular cylindrical model and the radiation patterns of THz transceivers were approximated using an idealistic antenna model.}, we investigate the coverage probability of downlink transmission in an indoor THzCom system.
We consider the joint effect of different types of blockages, directional antennas, and interference from nearby transmitters in a 3D THzCom  environment. 
The main contributions of this work are as follows:
\begin{itemize}
\item  We characterize the joint impact of blockages caused by the user itself, moving humans and wall blockers in a 3D THzCom environment. Also, we consider the effect of 3D directional antennas and derive the \emph{hitting probability}, which is defined as the probability of the signal corresponding to the main lobe of an interferer reaching a user. We show analytically that the hitting probability first increases and reaches a maximum and then decreases. This trend is not captured in the prior studies  at the THz band which considered 2D antenna models. Specifically, the hitting probability is overestimated in such studies.
\item We develop a tractable analytical framework, using stochastic geometry, to evaluate the coverage probability of the considered 3D THzCom system. Specifically, we derive an expression for the coverage probability, by characterizing the regions where dominant interferers (i.e., those can cause outage by themselves) can exist, and the average number of interferers that exist in these regions. We verify our analysis by comparing it to simulation results.
\item Our results show that the coverage  improvement brought by the increase in antenna directivity at transmitters is higher than that brought by the increase in antenna directivity at receivers.
     We also find that an increase in the density of human blockers slightly improves the coverage probability when the transmission link of interest is in LoS, but reduces the overall coverage probability.
     Finally, we show that the vertical heights of the THz devices profoundly impact the coverage probability in THzCom systems; therefore, ignoring its impact leads to a substantial underestimation of system reliability.
\end{itemize}

\begin{table}[t]
\vspace{-2.5mm}
\hspace{-5mm}
\caption{Summary of Main Mathematical Symbols.}\vspace{-12mm}
\begin{center}
\begin{tabular}{|c|l|l||l|l|}
\hline
\parbox[t]{1mm}{\multirow{11}{*}{\rotatebox[origin=c]{90}{\textbf{System Parameters}}}}
&\textbf{Symbol}&\textbf{Description}   & \textbf{Symbol}&\textbf{Description} \\
\cline{2-3}\cline{4-5}
&\multicolumn{2}{|c||}{\textbf{\textit{Network Model Parameters}}} & \multicolumn{2}{|c|}{\textbf{\textit{Antenna Model Parameters}} ($\Psi \in \{\textrm{A},\textrm{U}\}$)}  \\
\cline{2-5}
&$h_{\textrm{A}}$, $h_{\textrm{U}}$ & Height of APs and UEs    &  $G_{\Psi}^{\textrm{m}}$, $G_{\Psi}^{\textrm{s}}$&  Antenna gains of the main lobe and side lobes \\
&$\lambda_{\textrm{A}}$    &Density of APs & $\varphi_{\Psi,\textrm{H}}$, $\varphi_{\Psi,\textrm{V}}$   & Horizontal and vertical beamwidths of the antenna  \\
&$x_{ij}$, $d(x_{ij})$ & 2D and 3D distances of AP$_{i}$ to UE$_{j}$ link    &$\Omega_{\Psi}^{\textrm{m}}$,  $\Omega_{\Psi}^{\textrm{s}}$     &Solid angles of the main lobe and the side lobes  \\ \cline{2-3}
&\multicolumn{2}{|c||}{\textbf{\textit{Blockage Model Parameters}}}   & $k_{\Psi}$ & Coefficient of losses to side lobes, $k_{\Psi} \in (0,1)$  \\ \cline{2-5}
&$\omega$  & Self-blockage angle    &  \multicolumn{2}{c|}{\textbf{\textit{Propagation Model Parameters}}}    \\  \cline{4-5}
&$h_{\textrm{B}}$, $w_{1}$, $w_2$   & Height and widths of human blockers & $P_{\textrm{T}}$, $\sigma^{2}$  & Transmit power and AWGN power \\
&$v_{\textrm{B}}$    &Speed of human blockers & $f$      &Operating frequency  \\
&  $L_{\textrm{W}}$    & Length of wall blockers&  $K(f)$  & Absorption coefficient   \\
&$\lambda_{\textrm{B}}$, $\lambda_{\textrm{W}}$ &Densities of human and wall blockers &  $\tau$ & SINR threshold  \\  \hline
\parbox[t]{1mm}{\multirow{12}{*}{\rotatebox[origin=c]{90}{\textbf{Quantities of Interest}}}}
 &
\textbf{Symbol}  & \multicolumn{3}{|l|}{\textbf{Description} } \\ \cline{2-5}
&
$P_{\textrm{r}}^{\kappa,\iota}(x_{ij})$ & \multicolumn{3}{|l|}{Received power at UE$_{j}$ from AP$_{i}$  } \\
 &
$p_{\textrm{LoS,B}}(x_{ij})$  & \multicolumn{3}{|l|}{LoS probability for the link between AP$_{i}$ and UE$_{j}$ in the presence of only dynamic human blockers    } \\  &
$p_{\textrm{LoS,W}}(x_{ij})$ & \multicolumn{3}{|l|}{LoS probability for the link between AP$_{i}$ and UE$_{j}$ in the presence of only wall blockers   } \\  &
$p_{\textrm{LoS}}(x_{ij})$ & \multicolumn{3}{|l|}{LoS probability for the link between AP$_{i}$ and UE$_{j}$ in the presence of both human and wall blockers  } \\  &
$f_{x}(x_{ii})$ & \multicolumn{3}{|l|}{PDF of the horizontal distance between AP$_{i}$ and UE$_{i}$ in the presence of wall blockers     } \\  &
$p_{\textrm{hp}}(x_{i0})$ & \multicolumn{3}{|l|}{Hitting probability (Probability of the signal corresponding to the main lobe of AP$_{i}$ reaching  UE$_{0}$)    } \\  &
$p_{\textrm{hp,H}}(x_{i0})$ & \multicolumn{3}{|l|}{ Horizontal hitting probability    } \\  &
$p_{\textrm{hp,V}}(x_{i0})$ & \multicolumn{3}{|l|}{ Vertical hitting probability    } \\  &
$D_{\kappa,\iota}$ & \multicolumn{3}{|l|}{Distance from UE$_{0}$ to the boundary of the region around UE$_{0}$ where \emph{dominant interferers} can exist   } \\ &
$\Lambda_{\Phi^{\textrm{N}}}$, $\Lambda_{\Phi^{\textrm{F}}}$ & \multicolumn{3}{|l|}{Average number of \emph{near} and \emph{far dominant interferers} that exist around UE$_{0}$    } \\   &
$p_{\textrm{c}}(x_{00})$& \multicolumn{3}{|l|}{Coverage probability at UE$_{0}$    } \\  &
$ p_{\textrm{c},\textrm{LoS}}(x_{00})$ & \multicolumn{3}{|l|}{Coverage probability at UE$_{0}$ when the link between UE$_{0}$ and AP$_{0}$ is in LoS     } \\ \hline
\multicolumn{5}{l}{****$\textrm{A}$ is for AP, $\textrm{U}$ is for UE, $\textrm{B}$ is for human blocker, $\textrm{W}$ is for wall blocker, $\textrm{H}$ is for horizontal, $\textrm{V}$ is for vertical, $\textrm{m}$ is for }\\
\multicolumn{5}{l}{ main lobe, and $\textrm{s}$ is for side lobes.}
\end{tabular}\label{tab1}
\end{center}\vspace{-11mm}
\end{table}

The rest of the paper is organized as follows. In Section \ref{sec:system_model}, we describe the system model. In Section \ref{sec:PremResults}, we evaluate the  impact of blockages and 3D directional antennas  to obtain results that provide the foundation for the coverage analysis. In Section \ref{sec:Covanalysis}, we presents the analytical framework that is used to derive the coverage probability.  In Section \ref{sec:numerical}, numerical and simulation results are provided. Finally, in Section \ref{sec:conclusions} we conclude the paper.
The summary of the main mathematical symbols employed in this work is given in Table I.


\vspace{-2mm}
\section{System Model}\label{sec:system_model}

Fig.~\ref{Fig:SystemModel} depicts the 3D THzCom system
 considered in this work. We focus on the downlink signal propagation of a typical user who is at the center of a typical indoor environment.


\vspace{-3mm}
\subsection{Network Deployment}

 We consider that the THz access points (APs) are mounted on the ceiling. Hence, we model them as having fixed height $h_{\textrm{A}}$ and their locations follow a Poisson point process (PPP) in $\mathbb{R}^{2}$ with the density of $\lambda_{\textrm{A}}$. We also assume that user equipments (UEs), all of which are of fixed height $h_{\textrm{U}}$, are distributed uniformly within the circle with the radius $R_{\textrm{T}}$ centered at each AP. Although multiple UEs may exist in each circle, we assume that each AP in the system associates with a single UE and the link between the AP and its associated UE is not blocked by wall blockers.

For the purpose of analysis, we select a UE-AP pair, among the multiple UE-AP pairs, such that the UE in that pair is located at the center of the indoor environment \cite{StochGeo2011}.
We denote the UE and the AP in that pair by UE$_{0}$ and AP$_{0}$, respectively.
Also, we assume that all the UE-AP pairs share the same frequency channel; hence, apart from AP$_{0}$, all the other APs in the indoor environment of interest act as ``interferers'' to UE$_{0}$. We denote these interfering APs by  AP$_{i}$, where $i=1,2,\ldots$, and the UEs that associate with these APs by UE$_{j}$, where $j=1,2,\ldots$ with $i=j$ for UE and the AP in a specific UE-AP pair. Moreover, we denote $x_{ij}$ and $d(x_{ij})=\sqrt{\hbar^2 {+}x_{ij}^2}$ as the horizontal and 3D distances between AP$_{i}$ and UE$_{j}$, respectively, with $\hbar=h_{\textrm{A}}{-}h_{\textrm{U}}$.



\vspace{-3mm}
\subsection{Blockage Model}

In our system, we consider that the blockage of a UE-AP link is caused by (i) the user itself, referred to as self-blockage, or (ii) the dynamic human blockers,  or (iii) static wall blockers. 

\subsubsection{Self-Blockage}

\begin{figure}[!t]
\centering
\includegraphics[width=0.9\columnwidth]{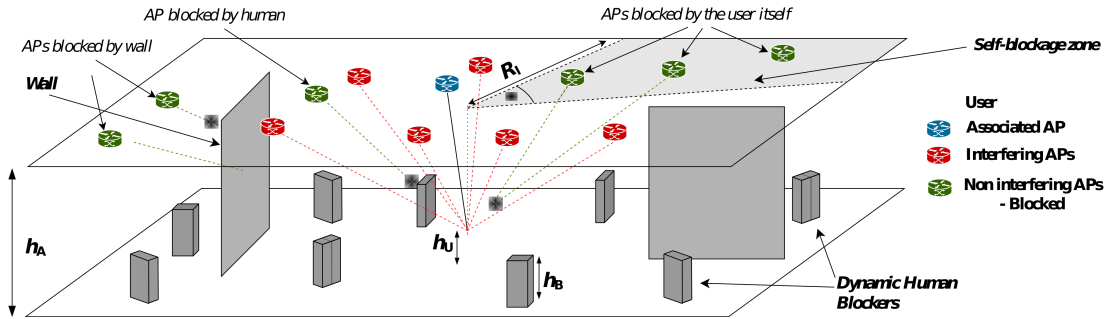}
\vspace{-7mm}
\caption{Illustration of the 3D THzCom system where a typical user associates with a non-blocked (blue) AP in the presence of interfering (red) APs. The non-interfering APs include those (green) blocked by self-blockage,  dynamic human and wall blockers.}\label{Fig:SystemModel}
\vspace{-8mm}
\end{figure}

Self-blockage plays a significant role in determining THz system performance. Notably, self-blockage may lead to the fact that signals from some APs surrounding a UE are totally blocked, even if the APs are within close proximity. We define the zone which is blocked by the UEs themselves as ``self-blockage zone'' \cite{mmO3,mmM2}, as shown in Fig.~\ref{Fig:SystemModel} with a self-blockage angle of $\omega$.

\subsubsection{Dynamic Human Blockers}

Humans moving in the area of the considered system can act as blockers. Specifically, they can potentially block the desired signals from AP$_{0}$ to UE$_{0}$, as well as the interference signals from other APs to UE$_{0}$.
We model the human blockers by rectangular absorbing screens (commonly referred to as the double knife-edge (DKE) model) with heights $h_{\textrm{B}}$  and widths $w_{1}$ and $w_2$  \cite{2016_VTC_mmWaveHumanBlockages_R3}, and their locations follow another PPP with the density of $\lambda_{\textrm{B}}$. Furthermore, we assume that the locations of humans form a PPP with the density of $\lambda_{\textrm{B}}$ and their mobility follows the random directional model (RDM)  \cite{RDM1}.
Based on the RDM model, a moving human randomly selects a direction in $(0,2\pi)$ to travel in and a duration for this travel. At the end of this duration, another independent direction to travel in and a time duration for this travel  are selected, and  the pattern continues. Following \cite{mmM2,MC1,MC2,MC3,akramICC2020}, in this work we assume that the moving speeds of all human blockers are $v_{\textrm{B}}$.

\enlargethispage{0.25 cm}

\subsubsection{Wall Blockers}
We employ a tractable Boolean scheme of straight lines to generate wall blockers in the indoor environment\footnote{The penetration losses of THz signals over wood, brick, plastic and glass materials are expected to be significantly higher due to the very small wavelength of THz signals \cite{2007_THzAttenuation_Kurner,2020_Kurner_2Vehicle}.
Thus, our proposed analysis can be extended even if glass or/and plywood walls are considered in the system model despite the fact that our analysis in this work primarily considers brick walls.}
\cite{IPN1,IPN1_Wall}. We assume that the lengths of walls, $L_{\textrm{W}}$, follow an arbitrary PDF of $f_{L_{\textrm{W}}}(L_{\textrm{W}})$ with mean $\mathbb{E}\left[L_{\textrm{W}}\right]$, and the centers of walls form a PPP of density $\lambda_{\textrm{W}}$. We ignore the widths of walls as the widths are much smaller compared to the lengths of walls.
Moreover, we assume that the orientations of walls are binary choice of either $0$ or $\pi/2$ with equal probability to ensure that walls are parallel or orthogonal to each other.  Furthermore, we assume that the heights of walls are fixed and are same as those of the APs, i.e., $h_{\textrm{W}}=h_{\textrm{A}}$.

\vspace{-2mm}
\subsection{Antenna Model}

\begin{figure}[!t]
  \centering
  \begin{minipage}[h]{0.37\columnwidth}\vspace{10mm}
   \includegraphics[width=0.9\columnwidth]{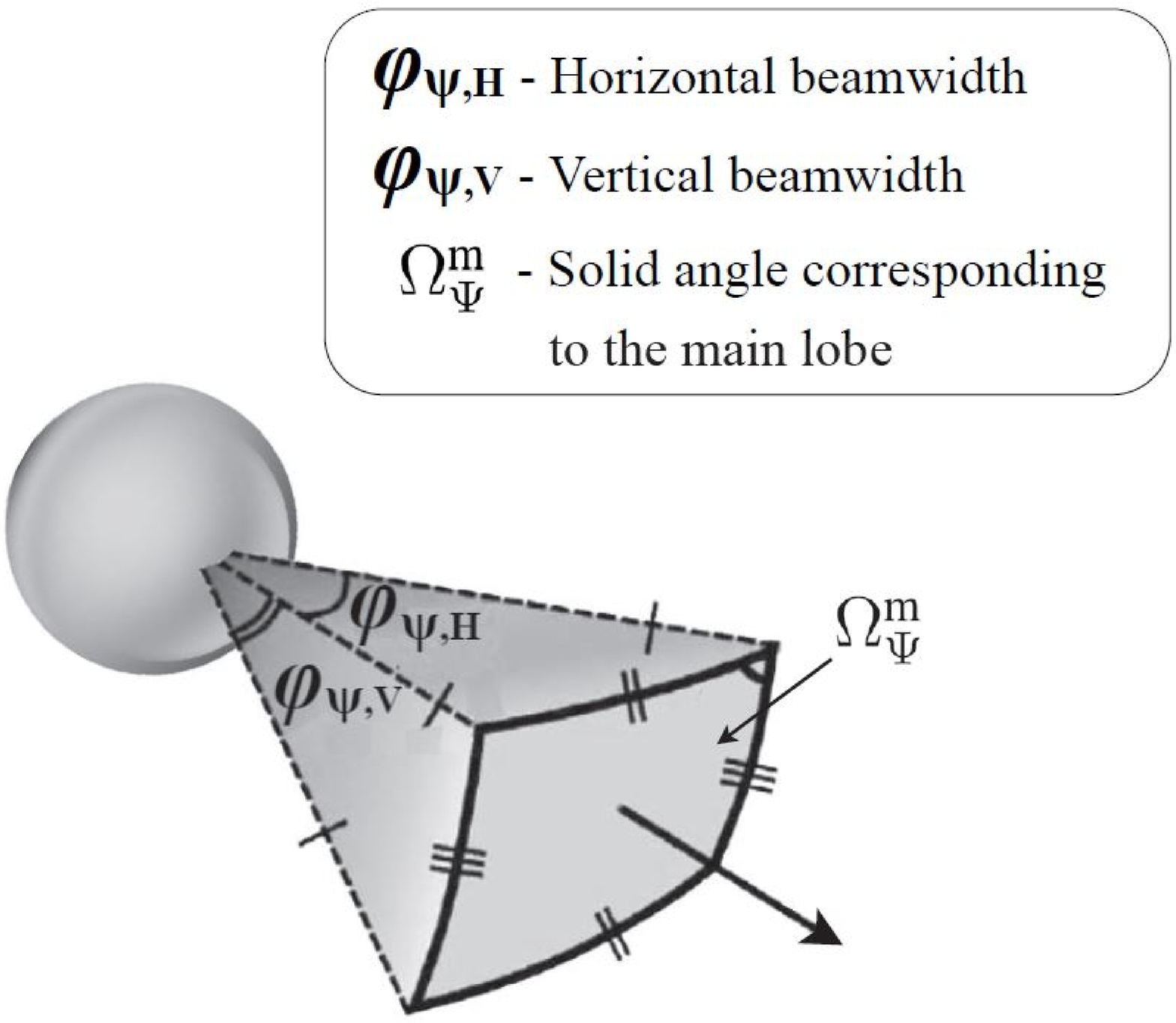}    \vspace{-6mm}
    \caption{3D sectored antenna radiation model.}
     \label{Fig:3Dbeam}
  \end{minipage}
  \hfill
  \begin{minipage}[h]{0.5\columnwidth}
    \includegraphics[width=\columnwidth]{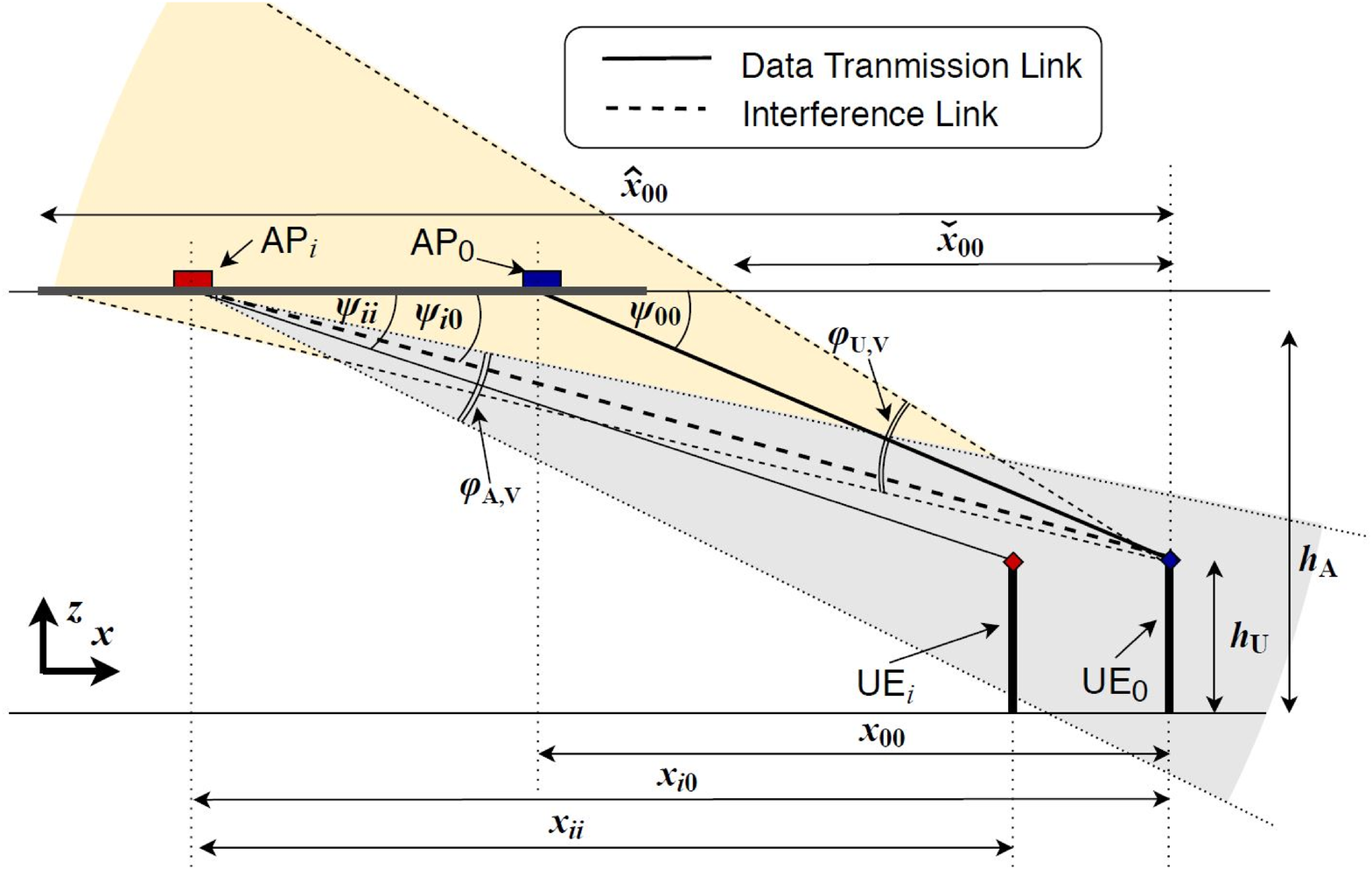}
   \vspace{-11mm}
    \caption{Illustration of the side view of the UE$_{0}$-AP$_{0}$ link.}
     \label{Fig:InfModel}
  \end{minipage}
        \vspace{-4mm}
\end{figure}

In this work, we assume that 3D beams are utilized at the APs and the UEs.
The 3D antenna beams are approximated by a 3D pyramidal-plus-sphere sectored antenna model, as shown in Fig.~\ref{Fig:3Dbeam} \cite{2019NN1,2017M5}. In this model, the pyramidal zone accounts for the main lobe of the antenna beam and the sphere accounts for the side lobes of the antenna beam \footnote{We clarify that sectored antenna models are widely used in the prior studies at the mmWave and THz bands for tractable analysis \cite{2020_ChongCoverage,2017M12,2020_WCL_Multi_RATMCforTHz,mmO3,2019NN1,Nan_ImprovingCoverageD2D,LaplaceIntRef2,2017M5,2018NN3,2017O1,2017N3}.
More importantly, the considered 3D sectored antenna model closely resembles the antenna radiation patterns that are observed in typical THz band horn antennas and graphene-based antennas \cite{2020_Kurner_2Vehicle,2017_ChongTVT_Graphene}.
However, the consideration of complex 3D antenna models such as the 3D sync model \cite{2017_PIMRC_SyncAntennaModel_R3} or the 3D multi-cone model \cite{2019GlobCom_Chong_Cov3DTHz} can further improve the accuracy of the results; thus will be considered in our future work.}. We also consider that the main lobe of APs and their associated UEs are tilted downwards and upwards, respectively, towards each other as shown in Fig. \ref{Fig:InfModel}. This guarantees beam alignment between APs and their associated UEs.

\enlargethispage{0.5 cm}

Based on the principles of antenna theory,  the antenna gains of the main lobe and the side lobes are expressed as
\begin{equation}
 \label{Equ:GMLProof1}
G_{\Psi}^{\textrm{m}}= \frac{P_{\textrm{rad,m}}/\Omega_{\Psi}^{\textrm{m}}}{P_{\textrm{rad}}/4\pi} ~~~ \textrm{and}~~G_{\Psi}^{\textrm{s}}= \frac{P_{\textrm{rad,s}}/\Omega_{\Psi}^{\textrm{s}}}{P_{\textrm{rad}}/4\pi},
\end{equation}
respectively, where $\Psi \in \{\textrm{A},\textrm{U}\}$ with $\textrm{A}$ is for AP and $\textrm{U}$ is for UE, $P_{\textrm{rad,m}}$ and $P_{\textrm{rad,s}}$ are the power concentrated along the main lobe and side lobes, respectively, with $P_{\textrm{rad}}=P_{\textrm{rad,m}}+P_{\textrm{rad,s}} $,  and $\Omega_{\Psi}^{\textrm{m}}$ and  $\Omega_{\Psi}^{\textrm{s}}$ are the solid angles corresponding to the main lobe and the side lobes, respectively.

Using the ``standard formula of rectangular plane'' given by the HCR's Theory of Polygon \cite{HCRPolygon1},
 $\Omega_{\Psi}^{\textrm{m}}$ in \eqref{Equ:GMLProof1} can be expressed as  $\Omega_{\Psi}^{\textrm{m}}=4 \arcsin\left(\tan\left(\frac{\varphi_{\Psi,\textrm{H}}}{2}\right) \tan\left(\frac{\varphi_{\Psi,\textrm{V}}}{2}\right)\right)$,
where $\varphi_{\Psi,\textrm{H}}$ and $\varphi_{\Psi,\textrm{V}}$ are the horizontal and vertical beamwidths of the antenna, respectively.
In addition, following the fact that the solid angle of the sphere is $4\pi$, we obtain $\Omega_{\Psi}^{\textrm{s}}$ as $\Omega_{\Psi}^{\textrm{s}}=4\pi - \Omega_{\Psi}^{\textrm{m}}$. Finally, by denoting $k_{\Psi}$ as the
 ratio of the fraction of power concentrated along the side lobes to the fraction of power concentrated along the main lobe, we obtain  the antenna gains of the main lobe and the side lobes in \eqref{Equ:GMLProof1} 
 as
\begin{equation}
\label{Equ:GML}
G_{\Psi}^{\textrm{m}}=  \frac{4 \pi}{\left(k_{\Psi}+1\right)\Omega_{\Psi}^{\textrm{m}}}  ~~~ \textrm{and}~~ G_{\Psi}^{\textrm{s}}= \frac{4 \pi k_{\Psi}}{\left(k_{\Psi}+1\right)\left(4\pi -\Omega_{\Psi}^{\textrm{m}}\right)},
\end{equation}
respectively.

\vspace{-3mm}
\subsection{Propagation Model}

The signal propagation at the THz band is determined by spreading loss and molecular absorption loss \cite{2011_Jornet_TWC}. Therefore, the received power at UE$_{j}$ from AP$_{i}$  in the 3D channel is\footnote{In this work, we utilize the far field propagation model since the transmission distances in the considered THzCom environment are very high compared to the propagation wavelength at the THz band.}
\begin{align}\label{Equ:Pr}
P_{\textrm{r}}^{\kappa,\iota}(x_{ij})
=&g_{\kappa,\iota}\;(d(x_{ij}))^{-2}e^{-K(f)d(x_{ij})},
\end{align}
where $g_{\kappa,\iota}\;\triangleq P_{\textrm{T}}G_{\textrm{A}}^{\kappa}G_{\textrm{U}}^{\iota}\left(c/4\pi f\right)^{2}$,
$P_{\textrm{T}}$ is the transmit power, $G_{\textrm{A}}^{\kappa}$ and $G_{\textrm{U}}^{\iota}$ are the effective antenna gains at AP$_{i}$ and UE$_{j}$, respectively, corresponding to link between AP$_{i}$ and UE$_{j}$ with $\kappa\in\{\textrm{m,s}\}$  and $\iota\in\{\textrm{m,s}\}$ where $\textrm{m}$ is for main lobe and \textrm{s} is for side lobes, $c$ is the speed of light, $f$ is the operating frequency, and $K(f)$
is the  molecular absorption coefficient of the transmission medium\footnote{Although the molecular absorption coefficient is frequency-dependent, its variation within the bandwidth of interest is relative small when THzCom systems operate within THz transmission windows \cite{HBM3}. Thus, similar to \cite{2020_WCL_Multi_RATMCforTHz,2015_VTC_THzIntNormalDist,2017M5,2018NN3,2020_ChongCoverage}, we assume that the molecular absorption coefficient remains unchanged within the bandwidth of interest.}.  Here, $\left(c/4\pi f d(x_{ij})\right)^2$ represents the spreading loss and $e^{-K(f)d(x_{ij})}$ represents molecular absorption loss. For a given pressure-temperature-humidity setting, $K(f)$ is obtained from
$K(f)=\frac{p}{p_{\textrm{STP}}}\frac{T_{\textrm{STP}}}{T}\sum_{i,g}Q^{i,g}\sigma^{i,g}(f)$, where $p$ and $T$ are the pressure and the temperature of the transmission environment, respectively, $p_{\textrm{STP}}$ and $T_{\textrm{STP}}$ are the standard pressure and temperature, respectively, and $Q^{i,g}$ and $\sigma^{i,g}(f)$ are the total number of molecules per unit volume and the absorption cross section for the isotopologue $i$ of gas $g$ at the frequency $f$, respectively \cite{2011_Jornet_TWC}. The values of $Q^{i,g}$ and $\sigma^{i,g}(f)$ are obtained from the HITRAN database \cite{2008_Hitran}.

We note that any surface that has roughness comparable to the wavelength of the electromagnetic wave introduces high  reflection, diffraction, and scattering losses.
Due to this and the fact that the wavelength of THz signals is extremely low, surfaces that are considered smooth at lower frequencies become rough at the THz band, thereby causing high reflection, diffraction, and scattering losses at the THz band  \cite{2017_ChongTVT_Graphene,HBM3}.
These  lead to (i) the reduced  number of significant NLoS rays that constitute a THz signal and (ii)  the substantial attenuation of significant NLoS rays as compared to the LoS ray.
Thus, similar to \cite{2017M12,2015_VTC_THzIntNormalDist,2017M5,2018NN3,2017N3,2020_WCL_Multi_RATMCforTHz}, we ignore the impact of NLoS rays and focus only on the LoS rays of THz signals. Also, we omit the impact of fading in the THz channel as  prior studies at the THz band \cite{2017M12,2020_WCL_Multi_RATMCforTHz,2015_VTC_THzIntNormalDist,CovTHz2019ICCC,2017M5,2018NN3,2017N3,2017O1}.


\vspace{-3mm}
\section{Impact of Blockages and 3D Directional Antennas}\label{sec:PremResults}

In this section, we analyze the impact of blockages and 3D directional antennas to obtain results that provide the foundation for coverage analysis in Section \ref{sec:Covanalysis}.

\vspace{-3mm}
\subsection{Impact of Blockages}

We first analyze the LoS probabilities due to each type of blockage acting alone, and present the results in the following lemmas.

\begin{figure}[!t]
\centering\subfloat[Side view\label{1a}]{ \includegraphics[clip,height=1.5in]{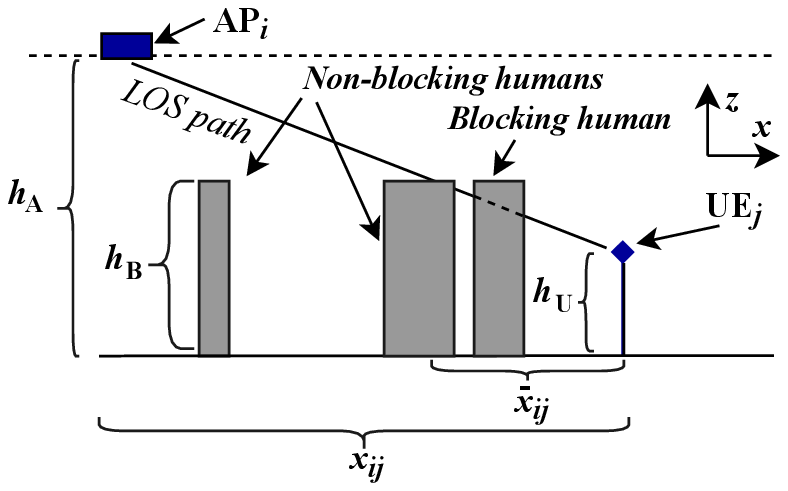}} \hspace{10 mm}
\subfloat[ {Top view}\label{1b}]{\includegraphics[clip,height=1.68in]{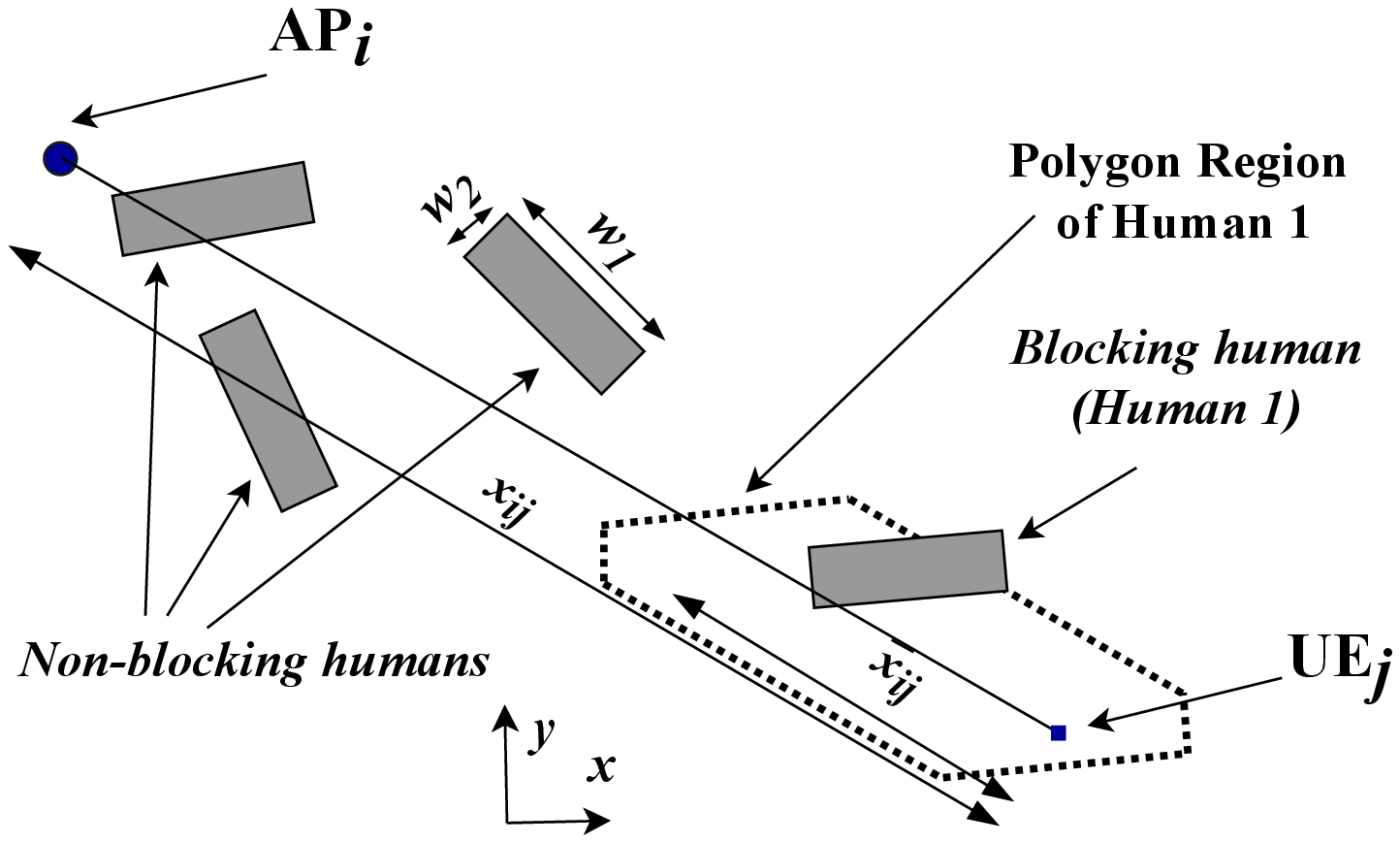}}
  \vspace{-1mm}
  \caption{Illustration of a single UE-AP link in the presence of dynamic human blockers.}\label{Fig:BlockModel}
  \vspace{-5mm}
\end{figure}

\begin{lemma} \label{Lem:pLH}
The LoS probability for the link between AP$_{i}$ and UE$_{j}$ in the presence of only dynamic human blockers is
\begin{equation}\label{Equ:pLH}
p_{\textrm{LoS,B}}(x_{ij})=\zeta e^{-\eta_{\textrm{B}} x_{ij}},
\end{equation}
where $\zeta = e^{-2 w_1 w_2 \lambda_{\textrm{B}}}$ and $\eta_{\textrm{B}} = \frac{2(w_1 + w_2)\lambda_{\textrm{B}}(h_{\textrm{B}}-h_{\textrm{U}})}{\pi \hbar}$.

\textit{~~~Proof:} According to the RDM model, if a human is moving in the area $\mathbb{R}^2$, the probability density function (PDF) of its location is uniform over time \cite{RDM1}. As such, at any given time instant, the locations of human blockers form a PPP with the same density of $\lambda_{\textrm{B}}$. This makes it possible to analyze a single time instant and then generalize the results over all time instants, thereby simplifying the mobility-related analysis.

Let us focus on a specific time instant. The region  in which a human blocker should appear to block the link  between AP$_{i}$ and UE$_{j}$ can be approximated by a polygon region between AP$_{i}$ and UE$_{j}$, as shown in Fig. \ref{Fig:BlockModel}(b). The dimensions of this polygon region depends on the orientation of the human blocker with respect to the UE$_j$-AP$_i$ link  \cite[Fig. 1]{mmM2ref11}.
Considering the area of this polygon region and the fact that the orientations of the human blockers are uniformly distributed between $0$ and $2 \pi$, we obtain the average number of human blockers that intersect the link between AP$_{i}$ and UE$_{j}$ in the 3D environment as
\begin{equation}\label{Equ:xbar}
\varpi_{ij}^{\textrm{B}}=\left(w_1 w_2 +\frac{2}{\pi }(w_1 + w_2)\bar{x}_{ij}\right)\lambda_{\textrm{B}},
\end{equation}
where $\bar{x}_{ij}=\frac{h_{\textrm{B}}-h_{\textrm{U}}}{\hbar}x_{ij}$.
Thereafter, considering the void probability of human blockers existing within the link between AP$_{i}$ and UE$_{j}$, we obtain $p_{\textrm{LoS,B}}(x_{ij})$ as
\begin{equation}\label{Equ:pLHProof1} 
p_{\textrm{LoS,B}}(x_{ij})=e^{-\varpi_{ij}^{\textrm{B}}}.
\end{equation}
Finally, substituting \eqref{Equ:xbar} in \eqref{Equ:pLHProof1}, we arrive at \eqref{Equ:pLH}.
 \hfill $\blacksquare$
\end{lemma}

\begin{lemma} \label{Lem:pLS}
The LoS probability for the link between AP$_{i}$ and UE$_{j}$ in the presence of only wall blockers  is
\begin{equation}\label{Equ:pLS}
p_{\textrm{LoS,W}}(x_{ij})=e^{-\eta_{\textrm{W}} x_{ij}},
\end{equation}
where $\eta_{\textrm{W}}=\lambda_{\textrm{W}}\frac{2}{\pi}\mathbb{E}\left[L_{\textrm{W}}\right]$.

\textit{~~~Proof:}
See Appendix \ref{app:Derive_pLS}.
\hfill $\blacksquare$
\end{lemma}

We clarify that in this work, we assume that each AP in the system selects its associating UE such that the link between the AP and its associated UE is not blocked by wall blockers. Thus, $p_{\textrm{LoS,W}}(x_{ii})=1,\forall i$.

We next jointly consider Lemma \ref{Lem:pLH} and Lemma \ref{Lem:pLS} to obtain the following corollary.

\begin{corollary} \label{Corol:pL}
The LoS probability for the link between AP$_{i}$ and UE$_{j}$ in the presence of both human and wall blockers  is
\begin{equation}\label{Equ:pL}
p_{\textrm{LoS}}(x_{ij})= p_{\textrm{LoS,B}}(x_{ij})p_{\textrm{LoS,W}}(x_{ij}) =\zeta e^{-\eta x_{ij}},
\end{equation}
where $\eta=\eta_{\textrm{B}}+\eta_{\textrm{W}}$.
\end{corollary}

We clarify that Corollary 1 is obtained by considering that the locations of humans and wall blockers are independent of each other and this  consideration has been  widely adopted in prior relevant studies in the literature for tractable analysis\footnote{In a realistic indoor environment, humans may keep a certain distance from walls. Thus, the consideration of dependency between the locations of humans and walls can improve the accuracy of the results and will be addressed in future work.} \cite{mmM2,2020_TVT_Blockage,MC1}.

\vspace{-3mm}

\subsection{Impact of 3D Directional Antennas}

To investigate the impact of an interfering AP on the  aggregated interference at UE$_{0}$, it is essential to characterize whether the signal corresponding to the main lobe of the interfering AP reaches UE$_{0}$. To this end, we determine the \emph{hitting probability}, $p_{\textrm{hp}}(x_{i0})$, which is defined as the probability of the signal corresponding to the main lobe of an interfering, i.e., AP$_{i}$ reaching  UE$_{0}$, where $i\neq 0$.

 The prior studies that analyze the coverage probability of THzCom systems approximated the antenna radiation pattern of THz transceivers using 2D antenna models  \cite{2017M12,2020_WCL_Multi_RATMCforTHz,2015_VTC_THzIntNormalDist,CovTHz2019ICCC,2017M5,2018NN3,2017N3,2017O1}.
In such studies, it was assumed that the main lobe of APs and their associated UEs are pointed towards each other in horizontal direction.
Under such consideration, the signal corresponding to the main lobe of the interfering AP reaches  UE$_{0}$, as long as UE$_{0}$ is within the horizontal beamwidths of the interfering AP, regardless of whether UE$_{0}$ is in close proximity to the interfering AP or not. Therefore, when a 2D antenna model is adopted,  the hitting probability is obtained as $p_{\textrm{hp}}(x_{i0})\vert_{\textrm{2D}}=\frac{\varphi_{\textrm{A,H}}x_{ij}}{2\pi x_{ij}}=\frac{\varphi_{\textrm{A,H}}}{2\pi}$ \cite{2020_WCL_Multi_RATMCforTHz}.

In contrast to the prior studies, in this work we assume that a 3D antenna model is utilized to approximate the antenna radiation patterns of transceivers.
Therefore, we consider that the main lobe of APs and their associated UEs are tilted downwards and upwards, respectively, towards each other as shown in Fig. \ref{Fig:InfModel}. This guarantees beam alignment between APs and their associated UEs.
Under such consideration,  the signal corresponding to the main lobe of an interfering AP reaches  UE$_{0}$ only when UE$_{0}$ is within both the horizontal and vertical beamwidths of the interfering AP.
Therefore, when the 3D sectored antenna model is adopted, $p_{\textrm{hp}}(x_{i0})$ is expressed as
\begin{equation} \label{Equ:pH}
    p_{\textrm{hp}}(x_{i0})=p_{\textrm{hp,H}}(x_{i0})p_{\textrm{hp,V}}(x_{i0}),
\end{equation}
where $p_{\textrm{hp,H}}(x_{i0})$ and $p_{\textrm{hp,V}}(x_{i0})$ are the horizontal and vertical hitting probabilities, respectively. $p_{\textrm{hp,H}}(x_{i0})$ and $p_{\textrm{hp,V}}(x_{i0})$ can be determined by evaluating the probabilities of UE$_{0}$ being within the horizontal and vertical beamwidths of AP$_{i}$, respectively \cite{2019NN1}.  Therefore,
$p_{\textrm{hp,H}}(x_{i0})$ is given by \vspace{-5mm}
\begin{align}\label{Equ:pHH} 
p_{\textrm{hp,H}}(x_{i0})=p_{\textrm{hp}}(x_{i0})\vert_{\textrm{2D}}=\frac{\varphi_{\textrm{A,H}}}{2\pi}.
\end{align}

We next derive $p_{\textrm{hp,V}}(x_{i0})$ in the following subsection.

\subsubsection{Vertical hitting Probability}

$p_{\textrm{hp,V}}(x_{i0})$ can be determined by evaluating the probabilities of UE$_{0}$ being within vertical beamwidths of AP$_{i}$. To determine this probability, it is essential to characterized the horizontal distance between AP$_{i}$ and UE$_{i}$ probabilistically.
Thus, we determine the PDF of the horizontal distance between AP$_{i}$ and UE$_{i}$  in the following lemma.

\begin{lemma} \label{Lem:PDFv}
The PDF of the horizontal distance between AP$_{i}$ and UE$_{i}$ in the presence of wall blockers is \vspace{-2mm}
\begin{equation}\label{Equ:PDFv}
f_{x}(x_{ii})=\begin{cases}
\varrho x_{ii} e^{-\eta_{\textrm{W}} x_{ii}}, &0\leq x_{ii}\leq R_{\textrm{T}},\\
0, &\textrm{otherwise},
\end{cases}
\end{equation}
where $\varrho=\eta^2_{\textrm{W}}/\left(1-e^{-\eta_{\textrm{W}}R_{\textrm{T}}}\left(1+\eta_{\textrm{W}}R_{\textrm{T}}\right)\right)
$ and
\begin{equation}\label{Equ:RT}
R_{\textrm{T}}=\sqrt{\left(\frac{2}{K(f)} W\left[\frac{K(f)}{2}\sqrt{\frac{g_{{\textrm{m},\textrm{m}}}\;}{\sigma^{2}\tau}}\right]\right)^2-\hbar^2}.
\end{equation}
Here, $\sigma^2$ is the additive white Gaussian noise (AWGN) power in the transmission window of interest, $\tau$ is the  predefined  SINR threshold, and $W\left[\cdot\right]$ is the Lambert \emph{W}-function.

\textit{~~~Proof:} See Appendix \ref{app:Derive_PDFv}. \hfill $\blacksquare$
\end{lemma}

Using  Lemma \ref{Lem:PDFv}, we next derive $p_{\textrm{hp,V}}(x_{i0})$ in the following proposition.

\begin{proposition} \label{Propos:pHV}
The vertical hitting probability for the link between AP$_{i}$ and UE$_{0}$ of the THzCom system in the typical indoor environment is
\begin{align}\label{Equ:pHV}
&p_{\textrm{hp,V}}(x_{i0})\notag \\
&=\begin{cases}
\frac{\varrho}{\eta_{\textrm{W}}^{2}}\Big[e^{-\eta_{\textrm{W}}\hbar\cot\left(\psi_{i0}+\frac{\varphi_{\textrm{A,V}}}{2}\right)
}\left(1+\eta_{\textrm{W}}\hbar\cot\left(\psi_{i0}+\frac{\varphi_{\textrm{A,V}}}{2}\right)\right)
&\\
~~~~~~~~~~~~~~~~-e^{-\eta_{\textrm{W}}\hbar\cot\left(\psi_{i0}-\frac{\varphi_{\textrm{A,V}}}{2}\right)
}\left(1+\eta_{\textrm{W}}\hbar\cot\left(\psi_{i0}{-}\frac{\varphi_{\textrm{A,V}}}{2}\right)\right)
\Big], &\!\! 0\leq x_{i0}\leq x_{\mu},\\
\frac{\varrho}{\eta_{\textrm{W}}^{2}}\!\left[e^{{-}\eta_{\textrm{W}}\hbar\!\cot\left(\psi_{i0}{+}\frac{\varphi_{\textrm{A,V}}}{2}\right)
}\!\!\left(1{+}\eta_{\textrm{W}}\hbar\cot\left(\psi_{i0}{+}\frac{\varphi_{\textrm{A,V}}}{2}\right)\right)\!
{-}e^{{-}\eta_{\textrm{W}}R_{\textrm{T}}
}\!\left(1{+}\eta_{\textrm{W}}R_{\textrm{T}}\right)
\right],\!\!\!& x_{\mu}\!\!<x_{i0}\!<\!x_{\nu},\\
0, & x_{i0} \geq x_{\nu},\end{cases}
\end{align}
where $\psi_{i0} = \arctan\left(\hbar/x_{i0}\right)$, $x_{\mu}= \hbar\cot\left(\textrm{min}\left\{\frac{\pi}{2},\bar{\psi}{{+}}\frac{\varphi_{\textrm{A,V}}}{2}\right\} \right)$, and $x_{\nu}=\hbar\cot \left(\textrm{max}\left\{0,\bar{\psi}-\frac{\varphi_{\textrm{A,V}}}{2}\right\}\right)$ with $\bar{\psi} = \arctan\left(\hbar/R_{\textrm{T}}\right)$.

\textit{~~~Proof:}
 See Appendix \ref{app:Derive_PHV}. \hfill $\blacksquare$
 \end{proposition}

\begin{remark} \label{Rem:pHV2Dand3D}
Prior studies that analyzed the coverage probability of THzCom systems adopted a 2D model and ignored the impact of $p_{\textrm{hp,V}}(x_{i0})$ by assuming it to be $p_{\textrm{hp,V}}(x_{i0})=1$ \cite{2017M12,2020_WCL_Multi_RATMCforTHz,2015_VTC_THzIntNormalDist,CovTHz2019ICCC,2017M5,2018NN3,2017N3,2017O1}.
However, by examining \eqref{Equ:pHV}, we observe that
$p_{\textrm{hp,V}}(x_{i0})$ increases when $x_{i0}$ increases up to $x_{\mu}$, and thereafter starts to decrease.
Furthermore, $p_{\textrm{hp,V}}(x_{i0})\leq 1$ for $\forall x_{i0}$.
These insights reveal that ignoring the impact of
$p_{\textrm{hp,V}}(x_{i0})$ in THzCom systems would lead to an overestimation of the hitting probability.
This in turn overestimates the interference, thereby leading to an underestimation of the coverage probability if a 2D model is adopted for analysis.  This will be illustrated in the results in Section \ref{sec:numerical}.
\end{remark}

Based on the derived results in this section, we next derive the coverage probability at UE$_{0}$ in the following section.

\vspace{-2mm}
\section{Coverage Analysis}\label{sec:Covanalysis}


The coverage probability at UE$_{0}$, $p_{\textrm{c}}(x_{00})$,  is the probability that the SINR at UE$_{0}$ is larger than the predefined threshold $\tau$, i.e., $p_{\textrm{c}}(x_{00}) = \mathbb{P}\left[\textrm{SINR} \geq \tau\right]$.
In the considered THzCom environment where blockages exist, $p_{\textrm{c}}(x_{00})$  can be written as
\begin{equation}\label{Equ:pcorg}
p_{\textrm{c}}(x_{00}) = p_{\textrm{LoS,B}}(x_{00}) p_{\textrm{c},\textrm{LoS}}(x_{00}),
\end{equation}
where $p_{\textrm{LoS,B}}(x_{00})$ is the LoS probability calculated in Lemma \ref{Lem:pLH} and $p_{\textrm{c},\textrm{LoS}}(x_{00})$ is the probability of the SINR at UE$_{0}$ is larger than $\tau$ when the link between UE$_{0}$ and AP$_{0}$ is in LoS. 


Conventionally, the coverage probability of communication systems has been derived with the aid of Laplace transform-based analysis when 2D antenna models are utilized to approximate the antenna radiation patterns of transceivers \cite{LaplaceIntRef2,StochGeo2011,Salman2014Stoch,2020_WCL_Multi_RATMCforTHz}. In such studies, the coverage probability is obtained based on the moment generating function of the aggregate interference, $\mathcal {L}_{I^{T}_{\mathrm {agg}}\mid x_{00}}(s)$. However, in this work we model the THzCom system in a 3D environment, thus a 3D antenna model is adopted to approximate the antenna radiation patterns of transceivers.
Under this consideration, it is challenging to obtain a tractable expression for $\mathcal {L}_{I^{T}_{\mathrm {agg}}\mid x_{00}}(s)$ \cite{2020_WCL_Multi_RATMCforTHz}.
 Specifically, the non-linear expression for the vertical hitting probability derived in Proposition 1 and the LoS probability derived in Corollary 1 have to be considered when determining $\mathcal {L}_{I^{T}_{\mathrm {agg}}\mid x_{00}}(s)$, thereby making the expression for  $\mathcal {L}_{I^{T}_{\mathrm {agg}}\mid x_{00}}(s)$ intractable. Hence, we need to resort to approximation methods to analyze the coverage probability  of 3D THzCom systems.



In this work, we use the dominant interferer analysis to derive the coverage probability. In doing so, we partition the APs which contribute to the aggregated interference at UE$_{0}$ into two subsets: \emph{dominant} and \emph{non-dominant interferers}.
We define an interferer as a \emph{dominant interferer} if it causes outage at UE$_{0}$ when none of the other interferers contribute to the aggregated interference  \cite{Nan_ImprovingCoverageD2D}. Moreover, we define an interferer as a \emph{non-dominant interferer} if it cannot cause outage by itself.
Dominant interferer analysis assumes that the presence of any combination of \emph{non-dominant interferers} cannot lead to the outage\footnote{
It is reasonable to assume that any combination of \emph{non-dominant interferers} cannot lead to the outage in THzCom systems, since the aggregated interference from distant interferers is minimal in such systems due to the following reasons. First, the interference power from a distant interferer is very small due to the exponential power decay as a result of the molecular absorption loss. Second, the probability of distant interferers causing interference at UE$_{0}$ is very low, due to the use of 3D directional antennas at the UEs and the APs and the fact that the LoS blockage exponentially increases with distance. We will validate the feasibility of this assumption in Section \ref{sec:numerical}-B.}.
This assumption allows the coverage probability of THzCom systems to be calculated analytically.

By using the dominant interferer analysis, $ p_{c,\textrm{LoS}}(x_{00}) $  can be interpreted as the probability that no interferer is a \textit{dominant interferer}, when the link between UE$_{0}$ and AP$_{0}$ is LoS.
Mathematically, $ p_{c,\textrm{LoS}}(x_{00}) $ is written as $p_{\textrm{c},\textrm{LoS}}(x_{00})=\mathbb{P}\left[n(\Phi)=0\right]$,
where $\Phi$ is the sets that denotes the \emph{dominant interferers} that exist around UE$_{0}$.
In calculating $\mathbb{P}\left[n(\Phi)=0\right]$,  for analytical simplicity, we further categorize the \emph{dominant interferers} into \emph{near} and \emph{far dominant interferers}.

We define an interferer as a \emph{near dominant interferer} if it can cause outage by itself while having its main lobe or the side lobes facing UE$_{0}$. Differently, we define an interferer as a \emph{far dominant interferer} if it causes outage by itself, only when its main lobe is facing UE$_{0}$.
Therefore, $ p_{\textrm{c},\textrm{LoS}}(x_{00}) $ can be re-interpreted as the probability that no interferer is a \emph{near} or a \emph{far dominant interferer}, when the link between UE$_{0}$ and AP$_{0}$ is LoS.
Mathematically, it is written as  \vspace{-5mm}
\begin{align}\label{Equ:pcL2}
p_{\textrm{c},\textrm{LoS}}(x_{00})=\mathbb{P}\left[n(\Phi)=0\right]=\mathbb{P}\left[n(\Phi^{\textrm{N}})=0\right] \mathbb{P}\left[n(\Phi^{\textrm{F}})=0\right],
\end{align}where $\Phi^{\textrm{N}}$ and $\Phi^{\textrm{F}}$ are the sets that denote the \emph{near} and \emph{far dominant interferers} that exist around UE$_{0}$, respectively.
To  find out the expression for $\mathbb{P}\left[n(\Phi^{\textrm{N}})=0\right]$, we denote $\Lambda_{\Phi^{\textrm{N}}}$ as the average number of \emph{near dominant interferers} that exist around UE$_{0}$. Therefore, considering the  null probability of $\Lambda_{\Phi^{\textrm{N}}}$, we obtain
\begin{align}\label{Equ:PnphiN}
\mathbb{P}\left[n(\Phi^{\textrm{N}})=0\right]=e^{-\Lambda_{\Phi^{\textrm{N}}}}.
\end{align}
Similarly, we denote $\Lambda_{\Phi^{\textrm{F}}}$ as the average number of \emph{far dominant interferers} that exist around UE$_{0}$. Considering the  null probability of $\Lambda_{\Phi^{\textrm{F}}}$, we obtain
  \begin{align}\label{Equ:PnphiF}
\mathbb{P}\left[n(\Phi^{\textrm{F}})=0\right]=e^{-\Lambda_{\Phi^{\textrm{F}}}}.
\end{align}
Thereafter, we substitute \eqref{Equ:PnphiN} and \eqref{Equ:PnphiF} into \eqref{Equ:pcL2} to obtain
\begin{equation} \label{Equ:pcL3}
    p_{\textrm{c},\textrm{LoS}}(x_{00})=e^{-\Lambda_{\Phi^{\textrm{N}}}-\Lambda_{\Phi^{\textrm{F}}}}.
\end{equation}
Finally, by substituting \eqref{Equ:pLH} and \eqref{Equ:pcL3} into \eqref{Equ:pcorg}, we obtain the coverage probability at UE$_{0}$ for the considered THzCom system in the typical indoor environment, and that is presented in the following theorem.

\begin{theorem} \label{Thrm:pc}
The coverage probability at UE$_{0}$ for the THzCom system in the typical indoor environment is
\begin{equation} \label{Equ:pc}
p_{\textrm{c}}(x_{00})=\zeta e^{-\eta_{\textrm{B}} x_{00}-\Lambda_{\Phi^{\textrm{N}}}-\Lambda_{\Phi^{\textrm{F}}}},
\end{equation}
where $\zeta$ and $\eta_{\textrm{B}}$ are defined in Lemma \ref{Lem:pLH}, and $\Lambda_{\Phi^{\textrm{N}}}$ and $\Lambda_{\Phi^{\textrm{F}}}$ are the average number of \emph{near} and \emph{far dominant interferers} that exist around UE$_{0}$, respectively.
\end{theorem}

\begin{figure}[t]
\centering
\vspace{-5mm}
\hspace{10mm}
\includegraphics[height=2in]{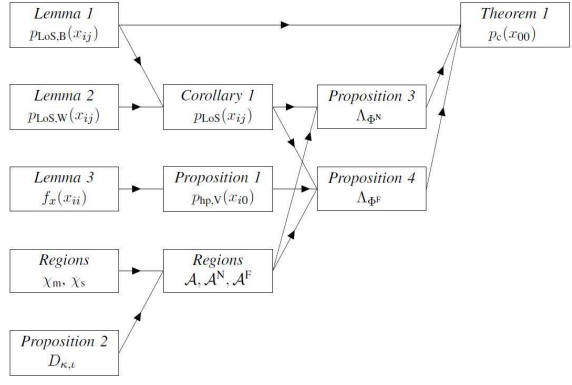}  \vspace{-6mm}
\caption{Summary of the analytical framework metrics.}\label{Fig:SummaryFig}
\vspace{-6mm}
\end{figure}

In Fig. \ref{Fig:SummaryFig}, we illustrate how the main results in this work are jointly utilized to obtain Theorem \ref{Thrm:pc}.
We next present the steps followed in obtaining the expression for $\Lambda_{\Phi^{\textrm{N}}}$  and $\Lambda_{\Phi^{\textrm{F}}}$  in the following subsection.

\vspace{-2mm}
\subsection{Derivation of $\Lambda_{\Phi^{\textrm{N}}}$ and $\Lambda_{\Phi^{\textrm{F}}}$}

\enlargethispage{0.5 cm}

As shown in Fig. \ref{Fig:SummaryFig}, to determine $\Lambda_{\Phi^{\textrm{N}}}$ and $\Lambda_{\Phi^{\textrm{F}}}$, it is essential to obtain $ \mathcal{A}$, where $\mathcal{A}$ is the region around UE$_{0}$ where \emph{dominant interferers} can exist. To obtain $\mathcal{A}$, we need to find (i) the region around UE$_{0}$ where the interfering APs exist that are within the main lobe and side lobes of UE$_{0}$, $\chi_{\textrm{m}}$ and $\chi_{\textrm{s}}$, respectively, and (ii) the distance from UE$_{0}$ to the boundary of the region around UE$_{0}$ where \emph{dominant interferers} can exist. We next find these quantities one after the other.

\begin{figure}[!t]
\centering
\subfloat[The regions $ \mathcal{A}_{\textrm{m,m}}$, $ \mathcal{A}_{\textrm{m,s}}$, $ \mathcal{A}_{\textrm{s,m}}$, and $ \mathcal{A}_{\textrm{s,s}}$ \label{2a}]{\includegraphics[clip,height=1.8in]{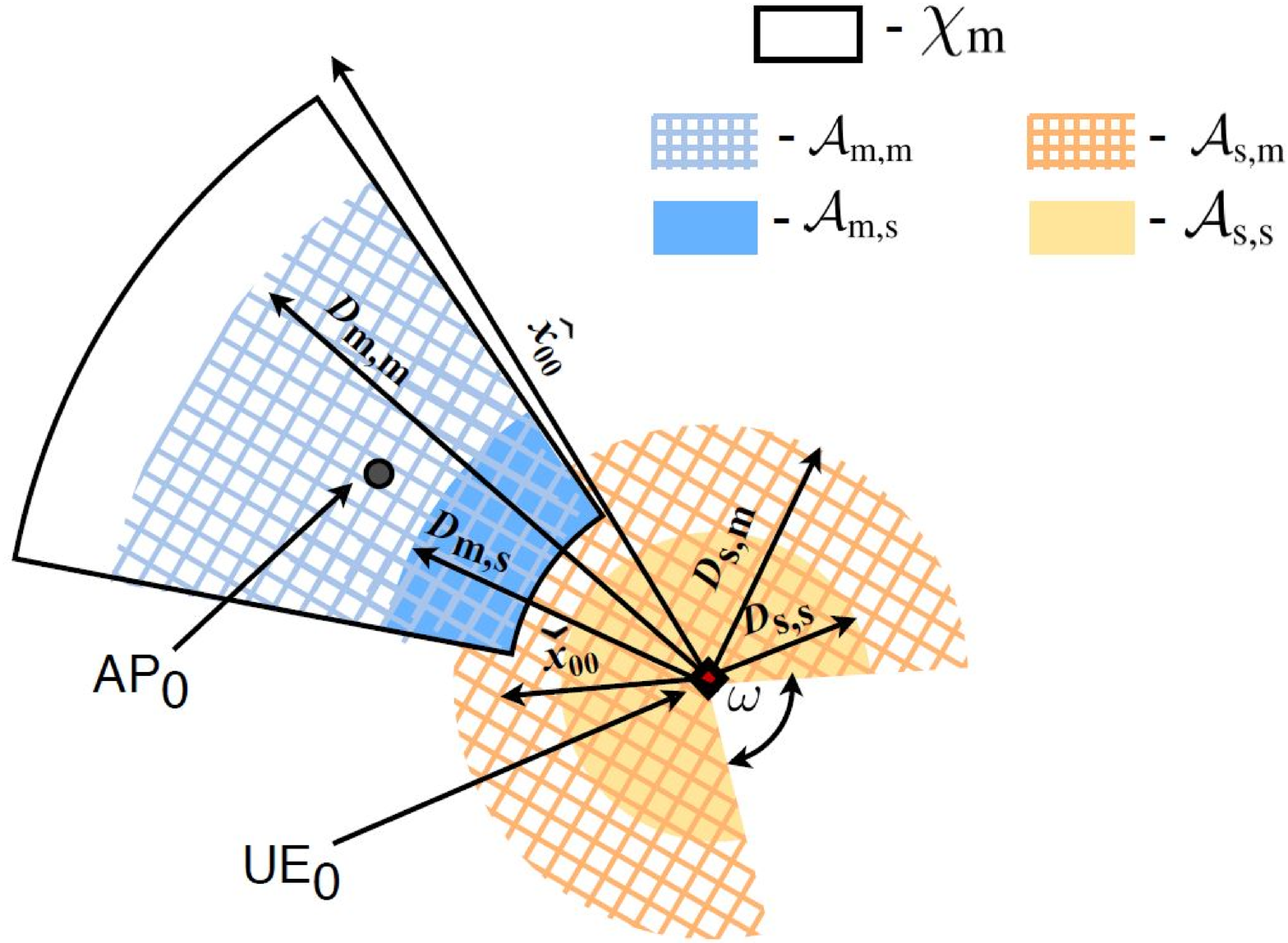}}\hfill
\subfloat[\emph{Near} and \emph{far dominant interferer} regions\label{2b}]{\includegraphics[clip,height=1.8in]{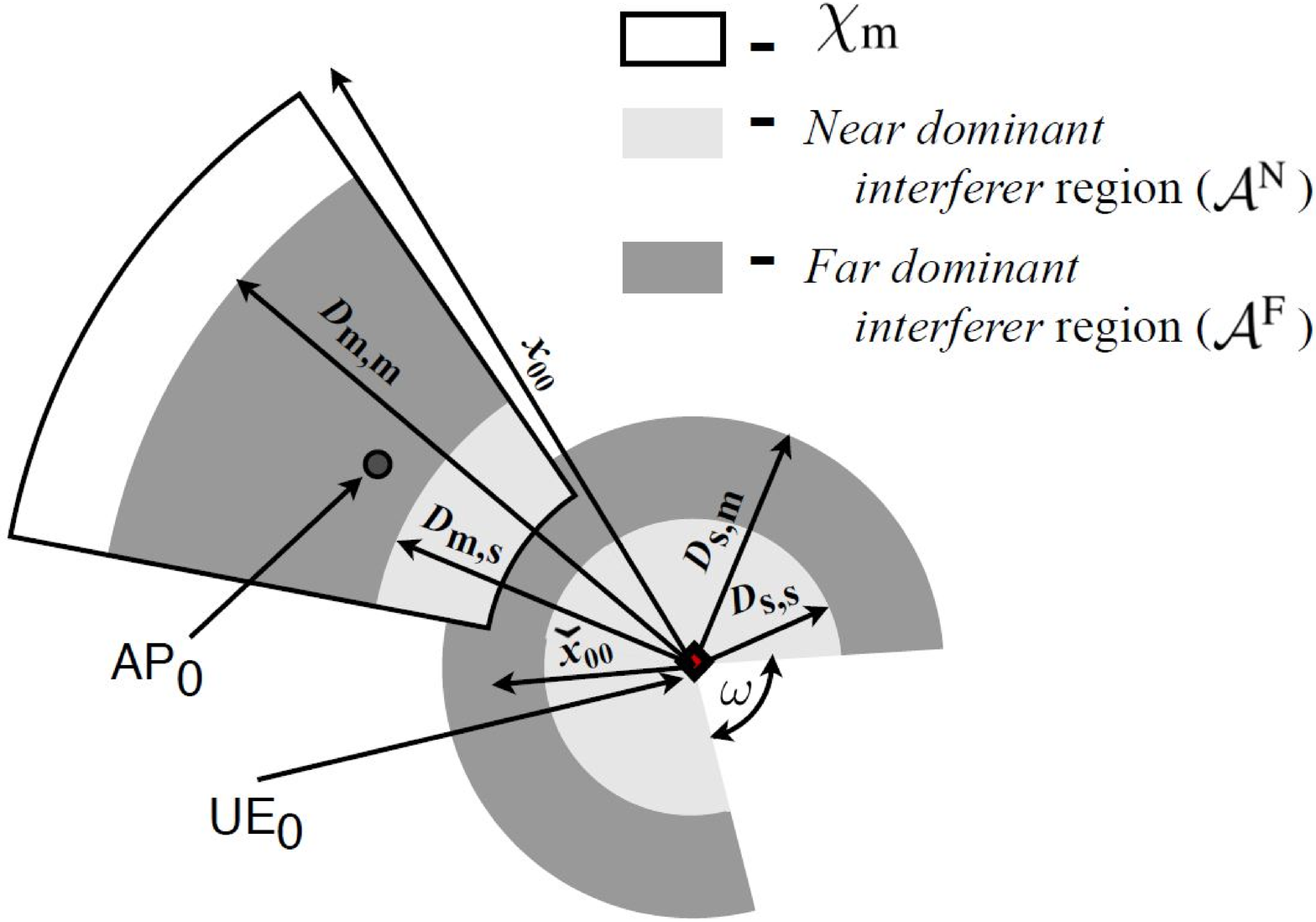}}
\caption{Illustration of the regions corresponding to the \emph{dominant interferers} when
$D_{\textrm{s},\textrm{s}}< \check{x}_{00} < D_{\textrm{s},\textrm{m}} < D_{\textrm{m},\textrm{s}} < D_{\textrm{m},\textrm{m}}< \hat{x}_{00}$.}\label{Fig:DomInt}
  \vspace{-5mm}
\end{figure}

Considering the vertical heights of the THz devices and the fact that directional antennas are used at UE$_{0}$,  we obtain $\chi_{\textrm{m}}$ as a truncated annular region in the horizontal plane, as shown in Fig.~\ref{Fig:DomInt}, where \vspace{-4mm}
\begin{equation}\label{Equ:chim}
\chi_{\textrm{m}}=\left\{(x,\theta),x\in\left[\check{x}_{00},\hat{x}_{00}\right],
\theta\in\Theta_{\textrm{m}}\right\},
\end{equation}
with \vspace{-5mm}
\begin{align}\label{Equ:rmin}
\check{x}_{00}=\begin{cases}
\frac{\hbar\left( x_{00}- \hbar \tan\left(\frac{\varphi_{\textrm{U,V}}}{2}\right)\right) }{\hbar+x_{00} \tan(\frac{\varphi_{\textrm{U,V}}}{2})}, &\textrm{if~}\psi_{00}\leq \frac{\pi-\varphi_{\textrm{U,V}}}{2},\\
0, &\textrm{otherwise},
\end{cases}
\end{align}\vspace{-3mm}
\begin{align}\label{Equ:rmax}
\hat{x}_{00}=\begin{cases}
\frac{\hbar\left( x_{00}+ \hbar \tan\left(\frac{\varphi_{\textrm{U,V}}}{2}\right)\right) }{\hbar-x_{00} \tan(\frac{\varphi_{\textrm{U,V}}}{2})}, &\textrm{if~}\psi_{00}\geq \frac{\varphi_{\textrm{U,V}}}{2},\\
\infty, &\textrm{otherwise},
\end{cases}
\end{align}
and , $\Theta_{\textrm{m}}=\left\{\theta, \theta\in \left[\theta_{00}-\frac{\varphi_{\textrm{U,H}}}{2},\theta_{00}
+\frac{\varphi_{\textrm{U,H}}}{2}\right]\right\}$.  We clarify that $\theta_{ij}$ is the angle that the projection of the UE$_{j}$-AP$_{i}$ link onto the horizontal plane forms with a given reference line in the horizontal plane.
Based on the knowledge of geometry and \cite[Eq (1.313.9)]{IntegralBook}, we obtain $\check{x}_{00}$ and $\hat{x}_{00}$ as in \eqref{Equ:rmin} and \eqref{Equ:rmax}, respectively \cite{akramICCWS2020}.
Next, considering the ``self-blockage zone'', we obtain $\chi_{\textrm{s}}$ as

\begin{equation} \label{Equ:chis}
\chi_{\textrm{s}}=\left\{(x,\theta), \left( x  \in \left(\left(0, \check{x}_{00}\right)\cup \left(\hat{x}_{00}, \infty\right) \right), \theta\in\Theta_{\textrm{m}} \right)  \cup  \left(x\in\left[0, \infty\right), \theta\in\Theta_{\textrm{s}}\right) \right\},
\end{equation}
where $\Theta_{\textrm{s}}=\left\{\theta, \theta \in \left(\left(\theta_{00}+\frac{\varphi_{\textrm{U,H}}}{2},\pi+ \theta_{00} - \frac{\omega}{2}\right] \cup \left[\pi+ \theta_{00} + \frac{\omega}{2}, 2\pi + \theta_{00}
-\frac{\varphi_{\textrm{U,H}}}{2}\right]\right)\right\}$.
Next, we derive the boundary of the region around UE$_{0}$ where \emph{dominant interferers} can exist in the following proposition.

\begin{proposition} \label{Propos:Dtaux0}
The distance from UE$_{0}$ to the boundary of the region around UE$_{0}$ where \emph{dominant interferers} can exist is
\begin{align} \label{Equ:Dtaux0}
D
=\sqrt{\left(\frac{2}{K(f)}W\left[\frac{K(f)}{2}\sqrt{\frac{g_{{\kappa,\iota}}\;\tau}
{P_{\textrm{r}}^{\textrm{m},\textrm{m}}(x_{00})-\tau\sigma^{2}}}\right]\right)^2-{\hbar}^2}.
\end{align}

\textit{~~~Proof:}
We recall that if an interferer is a \emph{dominant interferer},  it causes outage at UE$_0$ when none of the other interferers contribute to the aggregated interference. Therefore,  the SINR when only a \emph{dominant interferer} contributes to the aggregated interference should be less than the predefined SINR threshold $\tau$.
Mathematically, it is written as
\begin{equation}\label{Equ:Dtaux0Proof1}
  \frac{P_{\textrm{r}}^{\textrm{m},\textrm{m}}(x_{00})}{\sigma^{2}+g_{{\kappa,\iota}}\; (d(x_{i,0}^{D}))^{-2}e^{-K(f)d(x_{i,0}^{D})}} \leqslant  \tau .
  \end{equation}
where $x_{i,0}^{D}$ denotes the horizontal
distance between a \emph{dominant interferer} and UE$_0$. By rearranging \eqref{Equ:Dtaux0Proof1}, we obtain
\begin{equation}\label{Equ:Dtaux0Proof2}
\frac{K(f)d(x_{i,0}^{D})}{2}e^{\frac{K(f)d(x_{i,0}^{D})}{2}} \leqslant
\frac{K(f)}{2}\sqrt{\frac{g_{{\kappa,\iota}}\;\tau}{P_{\textrm{r}}^{\textrm{m},\textrm{m}}(x_{00})-\tau\sigma^{2}}}.
\end{equation}
Next, we apply the definition of the Lambert \textit{W}-function to \eqref{Equ:Dtaux0Proof2}, which leads to
\begin{equation}\label{Equ:Dtaux0Proof3}
d(x_{i,0}^{D})=
\sqrt{(x_{i,0}^{D})^2+\hbar^2} \leqslant \frac{2}{K(f)}W\left[\frac{K(f)}{2}\sqrt{\frac{g_{{\kappa,\iota}}\;\tau}
{P_{\textrm{r}}^{\textrm{m},\textrm{m}}-\tau\sigma^{2}}}\right] .
\end{equation}
Thereafter, we note that $D$ is given by $D=\max({x_{i,0}^{D})}$. Hence, we rearrange \eqref{Equ:Dtaux0Proof3} to obtain \eqref{Equ:Dtaux0}.
 \hfill $\blacksquare$
\end{proposition}

We clarify that there are four possibilities for $D$ in \eqref{Equ:Dtaux0}.
This is due to the fact that the effective antenna gains at the \emph{dominant interferer} and UE$_{0}$, corresponding to link between the \emph{dominant interferer} and UE$_{0}$, respectively, can each take two different values.
Considering this, we define the four possibilities for $D$ as $D_{\kappa,\iota}$ where $\kappa\in\{\textrm{m,s}\}$ and  $\iota \in\{\textrm{m,s}\}$.

We next jointly consider $\chi_{\textrm{m}}$, $\chi_{\textrm{s}}$, and $D_{\kappa,\iota}$ to obtain the region around UE$_{0}$ where \emph{dominant interferers} can exist, i.e., $\mathcal{A}$. We can obtain $\mathcal{A}$ as a combination of four regions which are denoted by $\mathcal{A}_{\kappa,\iota}$ where $\kappa\in\{\textrm{m},\textrm{s}\}$ and $\iota\in\{\textrm{m},\textrm{s}\}$.
For example, $\mathcal{A}_{\textrm{m,s}}$ denotes the region where \emph{dominant interferers} that are within the main lobe of UE$_{0}$ while having its side lobes facing UE$_{0}$ exist.
Following the fact that $D_{\kappa,\iota}\gtrless\check{x}_{00}$ and $D_{\kappa,\iota}\gtrless \hat{x}_{00}$, we obtain these region as
\begin{align} \label{Equ:chiD1}
 \mathcal{A}_{\textrm{m},\iota}=\left\{(x,\theta),x\in\left[\check{x}_{00},     v_{\textrm{m},\iota}\right],
\theta\in\Theta_{\textrm{m}}\right\},
\end{align}
where $\iota\in\{\textrm{m},\textrm{s}\}$, $v_{\textrm{m},\textrm{m}}=\min \{\hat{x}_{00}, D_{\textrm{m},\textrm{m}}\}$, and $v_{\textrm{m},\textrm{s}}=\max \{\check{x}_{00},\min \{\hat{x}_{00}, D_{\textrm{m},\textrm{s}}\}\}$, and
\begin{equation} \label{Equ:chiD3}
 \mathcal{A}_{\textrm{s},\iota}=\left\{(x,\theta), \left( x  \in \left(\left[0, v_{\textrm{s},\iota,1}\right]\cup \left[\hat{x}_{00},  v_{\textrm{s},\iota,2}\right] \right), \theta\in\Theta_{\textrm{m}} \right)  \cup  \left(x\in\left[0, D_{\textrm{s},\iota}\right], \theta\in\Theta_{\textrm{s}}\right) \right\},
\end{equation}
where  $v_{\textrm{s},\iota,1}=\min \{\check{x}_{00}, D_{\textrm{s},\iota}\}$ and $v_{\textrm{s},\iota,2}=\max \{\hat{x}_{00}, D_{\textrm{s},\iota}\}$ with $\iota\in\{\textrm{m},\textrm{s}\}$. Fig. \ref{Fig:DomInt}(a) illustrates these regions when
$D_{\textrm{s},\textrm{s}}< \check{x}_{00} < D_{\textrm{s},\textrm{m}} < D_{\textrm{m},\textrm{s}} < D_{\textrm{m},\textrm{m}}< \hat{x}_{00}$.
By examining these regions, we observe that $ \mathcal{A}_{\textrm{m,s}} \subset \mathcal{A}_{\textrm{m,m}} $ and $ \mathcal{A}_{\textrm{s,s}} \subset \mathcal{A}_{\textrm{s,m}} $. Hence, we can interpret that the interferers that are within the regions $ \mathcal{A}_{\textrm{m,s}}$ and $ \mathcal{A}_{\textrm{s,s}} $ cause outage by themselves while having its main lobe or the side lobes facing UE$_{0}$, i.e., \emph{near dominant interferers}. We denote the region where \emph{near dominant interferers} exist as $ \mathcal{A}^{\textrm{N}}$, where
\begin{align} \label{Equ:chiDN}
 \mathcal{A}^{\textrm{N}}&= \left( \mathcal{A}_{\textrm{m,m}} \cap \mathcal{A}_{\textrm{m,s}}\right) \cup \left( \mathcal{A}_{\textrm{s,m}} \cap \mathcal{A}_{\textrm{s,s}}\right) = \mathcal{A}_{\textrm{m,s}} \cup \mathcal{A}_{\textrm{s,s}}.
\end{align}
Similarly, we obtain the regions where the interferers that causes outage by itself, only when its main lobe is facing UE$_{0}$, i.e., \emph{far dominant interferers} exist as
\begin{align} \label{Equ:chiDF}
 \mathcal{A}^{\textrm{F}}&= \left( \mathcal{A}_{\textrm{m,m}} \cup \mathcal{A}_{\textrm{m,s}}\right) \cup \left( \mathcal{A}_{\textrm{s,m}} \cup \mathcal{A}_{\textrm{s,s}}\right) - \mathcal{A}^{\textrm{N}} = \mathcal{A}-\mathcal{A}^{\textrm{N}}.
\end{align}
Fig. \ref{Fig:DomInt}(b) illustrates the regions corresponding to the \emph{near} and \emph{far dominant interferers} when
$D_{\textrm{s},\textrm{s}}< \check{x}_{00} < D_{\textrm{s},\textrm{m}} < D_{\textrm{m},\textrm{s}} < D_{\textrm{m},\textrm{m}}< \hat{x}_{00}$.

Using the regions $\mathcal{A}^{\textrm{N}}$, $\mathcal{A}^{\textrm{F}}$ and the results in Section \ref{sec:PremResults}, we next derive the average number of \emph{near} and \emph{far dominant interferers} that exist around UE$_{0}$. The results are presented in the following propositions.

\begin{proposition} \label{Propos:densityN}
The average number of \emph{near dominant interferers} that exist around UE$_{0}$ for the THzCom system in the typical indoor environment is derived as
\begin{align}\label{Equ:densityN}
\!\!\!\Lambda_{\Phi^{\textrm{N}}}&=\frac{\lambda_{\textrm{A}} \zeta \varphi_{\textrm{U,H}}}{\eta^2} \Big[ {-}e^{{-}\eta v_{\textrm{s},\textrm{s},1}} (1{+} \eta v_{\textrm{s},\textrm{s},1}) {-}e^{{-}\eta v_{\textrm{m},\textrm{s}}} (1{+} \eta v_{\textrm{m},\textrm{s}})  {+} e^{{-}\eta \check{x}_{00}} (1{+} \eta \check{x}_{00}) {-}e^{{-}\eta v_{\textrm{s},\textrm{s},2}} (1{+} \eta v_{\textrm{s},\textrm{s},2}) \notag \!\!\!\!\!\! \\
&~~~   + e^{-\eta \hat{x}_{00}} (1+ \eta \hat{x}_{00})+e^{-\eta D_{\textrm{s},\textrm{s}}} (1+ \eta D_{\textrm{s},\textrm{s}})\Big]+ \frac{\lambda_{\textrm{A}} \zeta \left(2 \pi -\omega\right) }{\eta^2}  \left[1- e^{-\eta D_{\textrm{s},\textrm{s}}} (1+ \eta D_{\textrm{s},\textrm{s}}) \right].
\end{align}

\textit{~~~Proof:}
 See Appendix \ref{app:Derive_Prop234}. \hfill $\blacksquare$

\end{proposition}

\begin{proposition} \label{Propos:densityF}
The average number of \emph{far dominant interferers} that exist around UE$_{0}$ for the THzCom system in the typical indoor environment is derived as
\begin{align}\label{Equ:densityF}
\Lambda_{\Phi^{\textrm{F}}}&=\frac{\lambda_{\textrm{A}} \zeta \varphi_{\textrm{A,H}} \varphi_{\textrm{U,H}}}{2\pi} \Big[ \digamma(v_{\textrm{s},\textrm{s},1},v_{\textrm{s},\textrm{m},1}) +  \digamma(v_{\textrm{m},\textrm{s}},v_{\textrm{m},\textrm{m}})   +  \digamma(v_{\textrm{s},\textrm{s},2},v_{\textrm{s},\textrm{m},2}) -\digamma(D_{\textrm{s},\textrm{s}},D_{\textrm{s},\textrm{m}})\Big]\notag \\
& ~~~~~~~~~~~~~~~ + \frac{\lambda_{\textrm{A}} \zeta \varphi_{\textrm{A,H}} \left(2\pi-\omega\right)}{2\pi} \digamma(D_{\textrm{s},\textrm{s}},D_{\textrm{s},\textrm{m}}) ,
\end{align}
where $\digamma(a,b)=\int_{a}^{b} p_{\textrm{hp,V}}(x) e^{-\eta
 x} x \; dx$, which can be calculated numerically.

\textit{~~~Proof:}
 See Appendix \ref{app:Derive_Prop234}. \hfill $\blacksquare$

\end{proposition}

\vspace{-2mm}
\subsection{Special Case: Open Office Environment}

As mentioned in 3GPP standards, the open office environment where only human blockers exist is another interesting scenario for studies above-$6~\text{GHz}$ \cite{3GPPStand}.
Thus,  in this subsection we present the performance metrics of interest for the THzCom system in the open office environment.
Accordingly, in the following corollary, we first present the vertical hitting probability for the THzCom system in the open office environment.

\begin{corollary} \label{Corol:pHVOO}
The vertical hitting probability for the link between AP$_{i}$ and UE$_{j}$ of the THzCom system in the open office environment is
\begin{align}\label{Equ:pHVOO}
p_{\textrm{hp,V}}(x_{i0})=\begin{cases}
\frac{\hbar^{2}}{R_{\textrm{T}}^{2}}\left[\cot^2\left(\psi_{i0}-\frac{\varphi_{\textrm{A,V}}}{2}\right)
-\cot^2\left(\psi_{i0}+\frac{\varphi_{\textrm{A,V}}}{2}\right)\right], & 0\leq x_{i0}\leq x_{\mu},\\
1-\frac{\hbar^{2}}{R_{\textrm{T}}^{2}}\cot^{2}\left(\psi_{i0}+\frac{\varphi_{\textrm{A,V}}}{2}\right), & x_{\mu}<x_{i0}<x_{\nu},\\
0, & x_{i0} \geq x_{\nu}.\end{cases}
\end{align}


\textit{~~~Proof:} The proof is similar to that of Proposition \ref{Propos:pHV} and thus omitted. 
\hfill $\blacksquare$

\end{corollary}

We next present the  coverage probability at UE$_{0}$ for the THzCom system in the open office environment in the following theorem.

\begin{theorem} \label{Thrm:pcOO}
The coverage probability at UE$_{0}$ for the THzCom system in the open office environment is
\begin{equation} \label{Equ:pcOO}
p_{\textrm{c}}(x_{00})=\zeta e^{-\eta_{\textrm{B}} x_{00}-\Lambda_{\Phi^{\textrm{N}}_{\textrm{o}}}(x_{00})-\Lambda_{\Phi^{\textrm{F}}_{\textrm{o}}}(x_{00})},
\end{equation}
where $\Lambda_{\Phi^{\textrm{N}}_{\textrm{o}}}$ and $\Lambda_{\Phi^{\textrm{F}}_{\textrm{o}}}$ are the average number of \emph{near} and \emph{far dominant interferers} that exist around UE$_{0}$ for the THzCom system in the open office environment.
\end{theorem}

We clarify that  $\Lambda_{\Phi^{\textrm{N}}_{\textrm{o}}}(x_{00})$ can be obtained by replacing $\eta$ with $\eta_{\textrm{B}}$ in \eqref{Equ:densityN}.
Also, our derived result in \eqref{Equ:densityF} immediately leads to the following proposition, which delivers $\Lambda_{\Phi^{\textrm{F}}_{\textrm{o}}}$.

\begin{proposition} \label{Propos:densityFOO}
The average number of \emph{far dominant interferers} that exist around UE$_{0}$ for the THzCom system in the open office environment is derived as
\begin{align}\label{Equ:densityFOO}
\Lambda_{\Phi^{\textrm{F}}_{\textrm{o}}}(x_{00})&=\frac{\lambda_{\textrm{A}} \zeta \varphi_{\textrm{A,H}} \varphi_{\textrm{U,H}}}{2\pi} \Big[ \digamma_{\!\!\textrm{o}}(v_{\textrm{s},\textrm{s},1},v_{\textrm{s},\textrm{m},1}) +  \digamma_{\!\!\textrm{o}}(v_{\textrm{m},\textrm{s}},v_{\textrm{m},\textrm{m}})   +  \digamma_{\!\!\textrm{o}}(v_{\textrm{s},\textrm{s},2},v_{\textrm{s},\textrm{m},2}) -\digamma_{\!\!\textrm{o}}(D_{\textrm{s},\textrm{s}},D_{\textrm{s},\textrm{m}})\Big]\notag \\
&~~~~~~~~~~~~~~~~+ \frac{\lambda_{\textrm{A}} \zeta \varphi_{\textrm{A,H}} \left(2\pi -\omega\right)}{2\pi} \digamma_{\!\!\textrm{o}}(D_{\textrm{s},\textrm{s}},D_{\textrm{s},\textrm{m}}) ,
\end{align}
where \vspace{-2mm}
\begin{align}\label{Equ:F(a,b)}
\digamma_{\!\!\textrm{o}}(a,b)=\begin{cases}
I_{1}^{-}(b){-}I_{1}^{-}(a)+I_{1}^{+}(a){-}I_{1}^{+}(b), & 0\leq a \leq b  \leq  x_{\mu},\\
I_{1}^{-}(x_{\mu}){-}I_{1}^{-}(a)+I_{1}^{+}(a){-}I_{1}^{+}(b)+I_{2}(b){-}I_{2}(x_{\mu}), & 0 \leq a \leq x_{\mu}  \leq b  \leq x_{\nu},\\
I_{1}^{-}(x_{\mu}){-}I_{1}^{-}(a)+I_{1}^{+}(a){-}I_{1}^{+}(x_{\nu})+I_{2}(x_{\nu}){-}I_{2}(x_{\mu}), & 0 \leq a \leq x_{\mu},  b \geq x_{\nu}   ,\\
I_{2}(b){-}I_{2}(a)+I_{1}^{+}(a){-}I_{1}^{+}(b), &  x_{\mu} \leq a \leq b \leq x_{\nu}     ,\\
I_{2}(x_{\nu}){-}I_{2}(a)+I_{1}^{+}(a){-}I_{1}^{+}(x_{\nu}), &  x_{\mu} \leq a \leq  x_{\nu} \leq b     ,\\
0, &   a\geq   x_{\nu}.
\end{cases}
\end{align}
Here, \vspace{-1.5mm}
\begin{align}\label{Equ:I1pmab}
&I_{1}^{\pm}(x)=\begin{cases}
\frac{
\hbar^{2}}{2 R_{\textrm{T}}^{2}} \Bigg[ 2 e^{-\eta_{\textrm{B}} x} \cot(\frac{\varphi_{\textrm{A,V}}}{2}) \left(-\frac{ (1+\eta_{\textrm{B}} x) \cot(\frac{\varphi_{\textrm{A,V}}}{2}) }{\eta_{\textrm{B}}^2} \pm \frac{2\hbar \csc^2(\frac{\varphi_{\textrm{A,V}}}{2}) }{\eta_{\textrm{B}}} \pm \frac{\hbar^3 \csc^4(\frac{\varphi_{\textrm{A,V}}}{2}) }{x \pm \hbar \cot(\frac{\varphi_{\textrm{A,V}}}{2})} \right) \\
  ~~~~+ \hbar^2 e^{\pm \eta_{\textrm{B}} \hbar \cot(\frac{\varphi_{\textrm{A,V}}}{2})}  \csc^5(\frac{\varphi_{\textrm{A,V}}}{2}) \textrm{Ei}\left[- \eta_{\textrm{B}} x \mp \eta_{\textrm{B}} \hbar \cot(\frac{\varphi_{\textrm{A,V}}}{2}) \right]\\
  ~~~~~~~~ \times  \left(   \pm 2 \eta_{\textrm{B}} \hbar \cos(\frac{\varphi_{\textrm{A,V}}}{2}) +3 \sin(\frac{\varphi_{\textrm{A,V}}}{2}) +\sin(\frac{3\varphi_{\textrm{A,V}}}{2}) \right)\Bigg]+C_{1}^{\pm} ,&\!\!\!\!\!\!\!\!\eta_{\textrm{B}} \neq 0, \!\!\!\!\! \\
\frac{ \hbar^{2}}{R_{\textrm{T}}^{2}} \Bigg[ \frac{1}{2}\cot^2(\frac{\varphi_{\textrm{A,V}}}{2})x^2 \mp \hbar  \csc^4(\frac{\varphi_{\textrm{A,V}}}{2})\sin(\varphi_{\textrm{A,V}})x \pm \frac{\hbar^3 \csc^6(\frac{\varphi_{\textrm{A,V}}}{2}) \sin(\varphi_{\textrm{A,V}}) }{2 \left(x \pm \hbar \cot(\frac{\varphi_{\textrm{A,V}}}{2})\right)} \\
~  {+}\hbar^2 \left(2{+}\cos(\varphi_{\textrm{A,V}}) \right)\csc^4(\frac{\varphi_{\textrm{A,V}}}{2}) \ln\left(\hbar \cos(\frac{\varphi_{\textrm{A,V}}}{2}
)\pm x \sin(\frac{\varphi_{\textrm{A,V}}}{2}
)\right)   \Bigg] {+}C_{1}^{\pm}, &\!\!\!\!\!\!\!\!\!\text{otherwise},\!\!\!\!\!\!
\end{cases}
\end{align}
where $\textrm{Ei}\left[\cdot\right]$ is the exponential integral function and
\begin{align}\label{Equ:I2ab}
I_{2}(x)=\begin{cases}
  \frac{-(1+\eta_{\textrm{B}} x)e^{-\eta_{\textrm{B}} x}}{\eta_{\textrm{B}}^2}+C_{2}, &\eta_{\textrm{B}} \neq 0,\\
\frac{x^2}{2}+C_{2},&\text{otherwise},\!\!\!\!
\end{cases}
\end{align}
where $C_{1}^{\pm}$ and $C_{2}$ are the constants of the integrals.


\textit{~~~Proof:}  The proof is similar to that of Proposition \ref{Propos:densityF}
and thus omitted. It is noted that the steps followed in obtaining  $\digamma_{\!\!\textrm{o}}(a,b)$ in \eqref{Equ:F(a,b)} are presented in Appendix \ref{app:F(a,b)}. \hfill $\blacksquare$

\end{proposition}

\vspace{-1mm}
\section{Numerical Results and Discussion}\label{sec:numerical}
\vspace{-1mm}

In this section, we investigate the reliability performance of the considered 3D THzCom system. To this end, we first examine the impact of the derived hitting probability on the 3D THzCom system. Thereafter, we assess the accuracy and examine the significance of our derived coverage probabilities. Finally, we investigate the effects of system parameters on the coverage probabilities.
The simulation results are obtained using a framework that operates in a time-driven regime \cite{MC1} and considering a rectangular indoor environment of size $\ell_{1}\times \ell_{2}$ with $\ell_{1}=60~\textrm{m}$ and $\ell_{2}=50~\textrm{m}$ \cite{3GPPStand}. The values of the parameters used in this section are summarized in Table \ref{tab1}, unless specified otherwise. The chosen values are consistent with other relevant studies in the literature \cite{2011_Jornet_TWC,MC2,IPN1_Wall,mmM2,2020_Kurner_2Vehicle}.
In this work, we use the absorption coefficient values that are calculated for the standard atmosphere with $10\%$ humidity \cite{2011_Jornet_TWC}.  Also, we consider $\varphi_{\Psi,\textrm{H}}=\varphi_{\Psi,\textrm{V}}$.
While our analysis uses the dominant interferer assumption, the simulations consider the interference from all interferers that exist within the rectangular  indoor environment.

\vspace{-1mm}
\subsection{Hitting Probability}
\vspace{-1mm}

\begin{figure}[!t]
  \centering
  \begin{minipage}[h]{0.485\columnwidth}
   \includegraphics[width=\columnwidth]{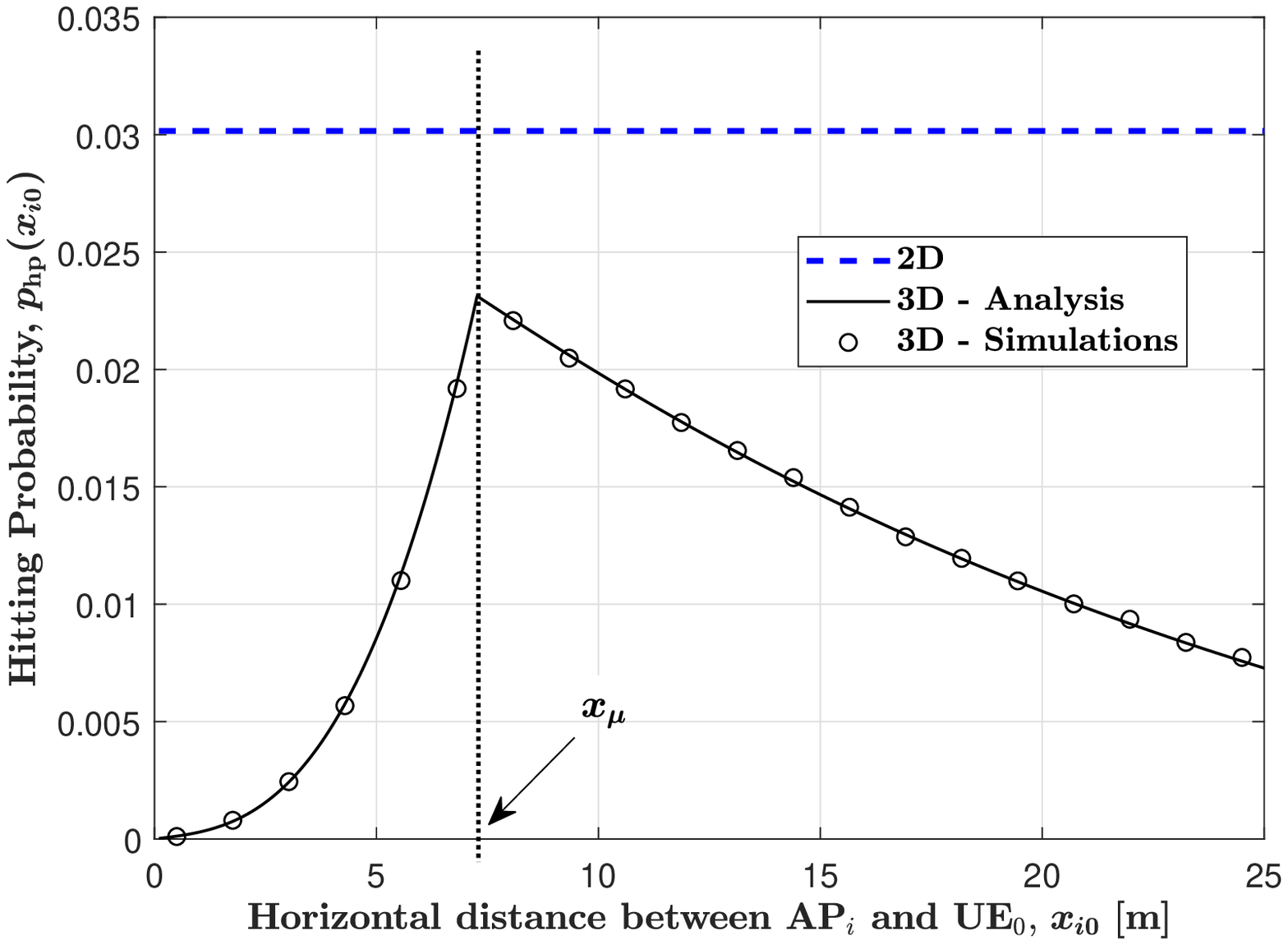}    \vspace{-11mm}
    \caption{Hitting probability versus the horizontal distance between UE$_{0}$ and AP$_{i}$ in the typical indoor environment for $R_{\textrm{T}}$=$12.2~\textrm{m}$.}
     \label{ResFig:pH1}
  \end{minipage}
  \hfill
  \begin{minipage}[h]{0.485\columnwidth}
    \includegraphics[width=\columnwidth]{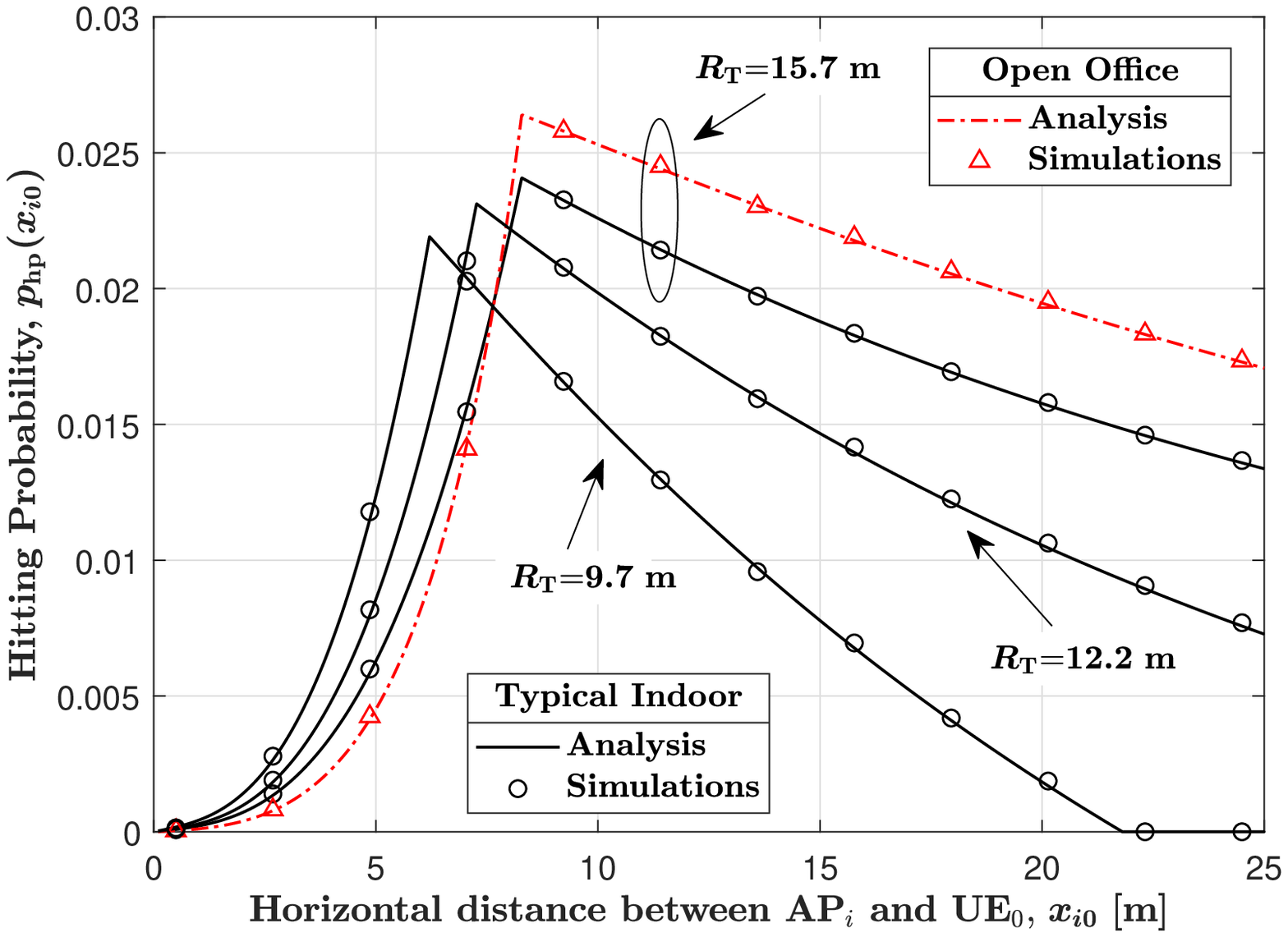}
   \vspace{-11mm}
    \caption{Hitting probability versus the horizontal distance between UE$_{0}$ and AP$_{i}$ in the open office environment, and in the typical indoor environment for different values of $R_{\textrm{T}}$.}
     \label{ResFig:pH2}
  \end{minipage}
        \vspace{-4mm}
\end{figure}

Fig.~\ref{ResFig:pH1} plots the hitting probability for the UE$_{0}$-AP$_{i}$ link, i.e., $p_{\textrm{hp}}(x_{i0})$ in Proposition \ref{Propos:pHV}, versus the horizontal distance between UE$_{0}$ and AP$_{i}$, $x_{i0}$, in the typical indoor environment for $R_{\textrm{T}}=12.2~\textrm{m}$.
In addition, the hitting probability for the 2D model also is plotted as a reference.
We first observe that the analytical results for the hitting probability match well with the simulation results, which demonstrates the correctness of our analysis.
Second, we observe that the hitting probability is very small for low $x_{i0}$, increases when $x_{i0}$ increases up to $x_{\mu}$, and thereafter starts to decrease.
This observation is accordance with Remark \ref{Rem:pHV2Dand3D}, which further validates our analysis.
Third, we observe that the hitting probability is significantly overestimated when the simplified 2D model is adopted as in prior studies that investigated coverage of THzCom systems. This demonstrates the importance of examining the performance of THz systems in 3D environment.

\begin{table}[t]
\caption{Value of System Parameters Used in Section~\ref{sec:numerical}}\vspace{-13mm}
\begin{center}
\begin{tabular}{|l|l|l||l|l|l|}
\hline
\textbf{Parameter} & \textbf{Symbol}& \textbf{Value}  & \textbf{Parameter} & \textbf{Symbol}& \textbf{Value}  \\
\hline
Height of APs and UEs & $h_{\textrm{A}}$, $h_{\textrm{U}}$  & $3.0~\textrm{m}$, $1.3~\textrm{m}$  & Transmit power  & $P_{\textrm{T}}$ &  $5~\textrm{dBm}$\\
\hline
Density of APs & $\lambda_{\textrm{A}}$   &  $0.1~\textrm{m}^{-2}$  & AWGN power  & $\sigma^{2}$ & $-77~\textrm{dBm}$ \\ \hline
Operating frequency  & $f$  & $1.05~\textrm{THz}$   &  SINR threshold& $\tau$ & $3~\textrm{dB}$  \\ \hline
Absorption coefficient   & $K(f)$ &  $0.07512~\textrm{m}^{-1}$   & Self-blockage angle&  $\omega$ & $60^{\circ} $   \\ \hline
Operating bandwidth  &  $B$  & $5~\textrm{GHz}$   & Height and density of human blockers & $h_{\textrm{B}}$, $\lambda_{\textrm{B}}$ & $1.7~\textrm{m}$,  $0.1~\textrm{m}^{-2}$    \\ \hline
AP's antenna &  $G_{\textrm{A}}^{\textrm{m}}$, $G_{\textrm{A}}^{\textrm{s}}$, $k_{\textrm{A}}$, &   $25~\textrm{dBi}$, ${-}10~\textrm{dBi}$,   &
Widths of human blockers &
$w_{1}$, $w_2$   &
$0.6~\textrm{m}$, $0.3~\textrm{m}$    \\\cline{4-6}
parameters&   $\varphi_{\textrm{A},\textrm{H}}$, $\varphi_{\textrm{A},\textrm{V}}$ &  $0.1$, $10^{\circ} $, $10^{\circ} $   & Speed of human blockers  & $v_{\textrm{B}}$  & $1~\textrm{ms}^{-1}$ \\ \hline
 UE's antenna & $G_{\textrm{U}}^{\textrm{m}}$, $G_{\textrm{U}}^{\textrm{s}}$, $k_{\textrm{U}}$, &  $15~\textrm{dBi}$, ${-}10~\textrm{dBi}$,   & Average length of wall blockers& $\mathbb{E}\left[L_{\textrm{W}}\right]$& $3.0~\textrm{m}$    \\\cline{4-6}
 parameters &  $\varphi_{\textrm{U},\textrm{H}}$, $\varphi_{\textrm{U},\textrm{V}}$   & $0.1$,  $33 ^{\circ} $, $33 ^{\circ} $   & Density of wall blockers & $\lambda_{\textrm{W}}$   &   $0.04~\textrm{m}^{-2}$     \\ \hline
\end{tabular}\label{tab1}
\end{center}\vspace{-9mm}
\end{table}

Fig.~\ref{ResFig:pH2} plots the hitting probability versus $x_{i0}$, for different values of $R_{\textrm{T}}$  in the typical indoor environment. Moreover, the hitting probability obtained in the open office environment is also plotted. We clarify that we obtain the plots of  the hitting probability  corresponding to different values of $R_{\textrm{T}}$ by changing $\tau$. Specifically, the values of $R_{\textrm{T}}$ of $15.7~\textrm{m}$, $12.2~\textrm{m}$, and $9.7~\textrm{m}$ correspond to the values of $\tau$ of $0~\textrm{dB}$, $3~\textrm{dB}$, and $6~\textrm{dB}$, respectively. We first observe that as compared to the typical indoor environment, for the open office environment the hitting probability is lower for low $x_{i0}$ and higher for large $x_{i0}$. This is due to the fact that in the typical indoor environment, APs associate with UEs that are not blocked by wall blockers. Therefore, it is more likely for AP$_{i}$ to associate with a farther UE in the open office environment as compared to the typical indoor environment. This in turn increases the probability of the UE associated with AP$_{0}$ being closer to UE$_{0}$, thereby increasing the hitting probability  for large $x_{i0}$. Second, we observe that when $R_{\textrm{T}}$ decreases, the hitting probability increases for low $x_{i0}$, and reduces for large $x_{i0}$.  These observations are expected, since the likelihood of AP$_{i}$ being associated with a closer UE increases when $R_{\textrm{T}}$ reduces. Finally, we observe that for some cases of large $x_{i0}$ and small $R_{\textrm{T}}$, e.g., $x_{i0}\geq 21.8~\textrm{m}$ when $R_{\textrm{T}}=9.7~\textrm{m}$,  the hitting probability is zero. This further indicates the importance  of examining the performance of THzCom systems in 3D environment.

\vspace{-1mm}
\subsection{Coverage Probability}
\vspace{-1mm}

\begin{figure}[t]
  \centering
  \begin{minipage}[h]{0.485\columnwidth}
   \includegraphics[width=\columnwidth]{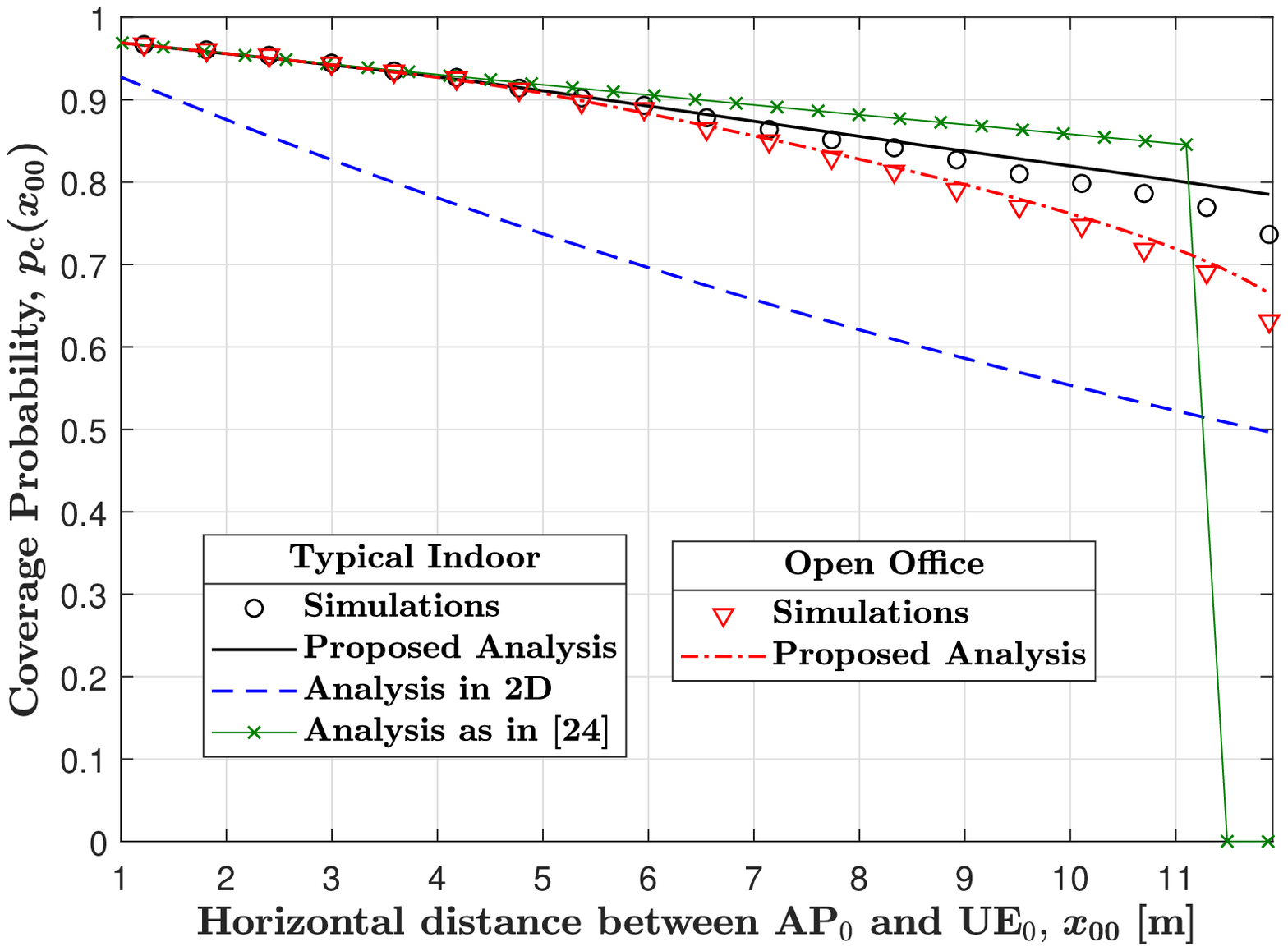}  \vspace{-12mm}
    \caption{The coverage probability versus the horizontal distance between AP$_{0}$ and UE$_{0}$ for the typical indoor and the open office environment.}
     \label{ResFig:Cova}
  \end{minipage}
  \hfill
  \begin{minipage}[h]{0.485\columnwidth}
   \includegraphics[width=\columnwidth]{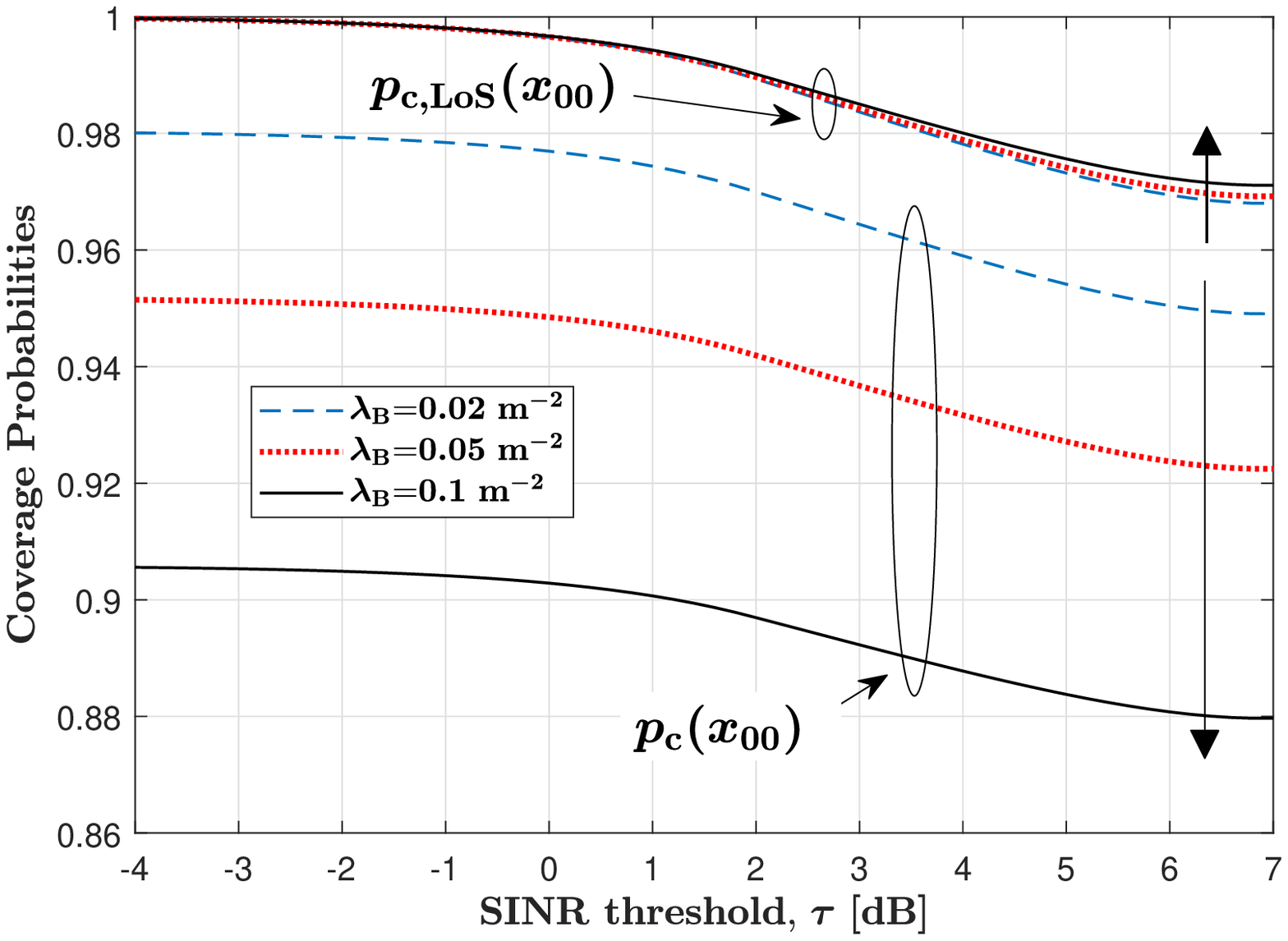}
   \vspace{-12mm}
    \caption{The coverage probability versus the SINR threshold for the typical indoor environment with different human blocker densities.}
    \label{ResFig:Covd}
  \end{minipage}
  \vspace{-5mm}
\end{figure}

Fig. \ref{ResFig:Cova}  plots the coverage probability versus the horizontal distance between AP$_{0}$ and UE$_{0}$ for the typical indoor and the open office environments.
In this figure, for the typical indoor environment, we plot (i) the simulated coverage probabilities in the considered in the 3D model, (ii) the coverage probabilities derived as per the proposed analysis  in the 3D model, (iii) the coverage probabilities derived as per the analysis in \cite{2017O1} in the 3D model, and (iv) the coverage probabilities derived as per the proposed analysis in the 2D model, which is obtained by setting $\bar{x}_{ij}=x_{ij}+r_{\textrm{B}}$ in \eqref{Equ:xbar} and $p_{\textrm{hp,V}}(x_{ij})=1$ for $\forall i,j$.

We first observe that our analysis well matches the simulations for small and medium $x_{00}$ for both the open office and the typical indoor environment, which demonstrates the correctness of our analysis.
For high  $x_{00}$, our analysis  slightly overestimates the coverage probability for both the open office and the typical indoor environment.
The slight overestimation for the open office environment is due to the fact that our analysis is under the assumption that any combination of \emph{non-dominant interferers} cannot lead to the outage.
 However, non-negligible possibilities of \emph{non-dominant interferers} causing outage  appear since there are more \emph{non-dominant interferers} within the beamwidth of the UE for high $x_{00}$, which yields the slight overestimation. Differently, for the typical indoor environment, the much smaller overestimation appears. This is because, although we consider that  the number of walls that intersect each link is independent, non-negligible dependencies appear in the number of walls that intersect each link.

Second, we observe a  gradual decrease in the coverage probability when $x_{00}$ increases for both the open office and the typical indoor environment. This is due to the fact that when the UE$_{0}$ is connected to a farther AP, in addition to the reduced received power, the impact of interference on the coverage probability becomes more detrimental since there are more interferers within the beamwidth of the UE$_{0}$. Moreover, the effective number of blockers that exist in the UE$_{0}$-AP$_{0}$ link increases with $x_{00}$, thereby further deteriorating the coverage probability. In addition, we observe that the coverage probability for the typical indoor environment is higher than that of the open office environment, especially when $x_{00}$ is high. This is due to the fact that since there are wall blockers in addition to human blockers in typical indoor environment, the likelihood of interference signals being blocked becomes higher, which improves the coverage probability.

We next compare the coverage probability obtained using our analysis with that obtained in prior studies.
We first observe that the coverage probability is significantly underestimated when the analysis is performed using the 2D model as in \cite{2017M12,2020_WCL_Multi_RATMCforTHz,2015_VTC_THzIntNormalDist,CovTHz2019ICCC,2017M5,2018NN3,2017N3,2017O1}.  This is due to the fact that in the 2D model the blockage and hitting probabilities are overestimated, since vertical heights of the THz devices are ignored.
This in turn underestimates the received power, and overestimates the interference at UE$_{0}$, thereby leading to  the  underestimation of the coverage probability.
This observation reveals that the vertical heights of THz devices profoundly impact the coverage probability of THzCom systems; therefore, ignoring them leads to an underestimation of the system reliability.
Moreover, we observe that the underestimation of the coverage probability when the analysis is performed using the 2D model, increases when the AP$_{0}$ to UE$_{0}$ link distance increases.Second, we observe that the analysis in \cite{2017O1}, which is based on average interference, approximates the coverage probability well for low $x_{00}$, but the accuracy significantly deteriorates when $x_{00}$ increases. In comparison, our analysis approximates the coverage probability very well for all $x_{00}$.

\subsubsection{\underline{Impact of SINR Threshold and Blocker Densities}}

Fig.~\ref{ResFig:Covd} plots the coverage probability, i.e, $p_{c}(x_{00})$ versus $\tau$, for different densities of human blockers, $\lambda_{\textrm{B}}$, when $x_{00}=6~\textrm{m}$.
We also plot the coverage probability when the UE$_{0}$-AP$_{0}$ link is in LoS, i.e., $p_{c,\textrm{L}}(x_{00})$ in \eqref{Equ:pcL3}, which is used as a metric of interest in \cite{MC3,akramICC2020,2020_ChongCoverage}.
We observe that $p_{c}(x_{00})$ and $p_{c,\textrm{L}}(x_{00})$ become lower when $\tau$ increases.
In addition, we observe that an increase in $\lambda_{\textrm{B}}$ leads to a slight improvement in $p_{c,\textrm{L}}(x_{0})$, but a significantly decrease in $p_{c}(x_{00})$.
When there are more blockers, the likelihood of interference signals being blocked becomes higher, which leads to higher $p_{c,\textrm{L}}(x_{0})$. However, the increase in $\lambda_{\textrm{B}}$ increases the likelihood of AP$_{0}$ being blocked, leading to worse $p_{c}(x_{00})$.
These observations on $p_{c}(x_{00})$ and $p_{c,\textrm{L}}(x_{0})$ for varying blocker densities indicate that it is important to carefully select system parameters, e.g., antenna gains, transmit power or density of APs,  to obtain the desired reliability performance depending on the density of humans in the indoor environment.


\subsubsection{\underline{Impact of  Antenna Parameters of UEs and APs}}

\begin{figure}[t]
  \centering
  \begin{minipage}[h]{0.485\columnwidth}
   \includegraphics[width=\columnwidth]{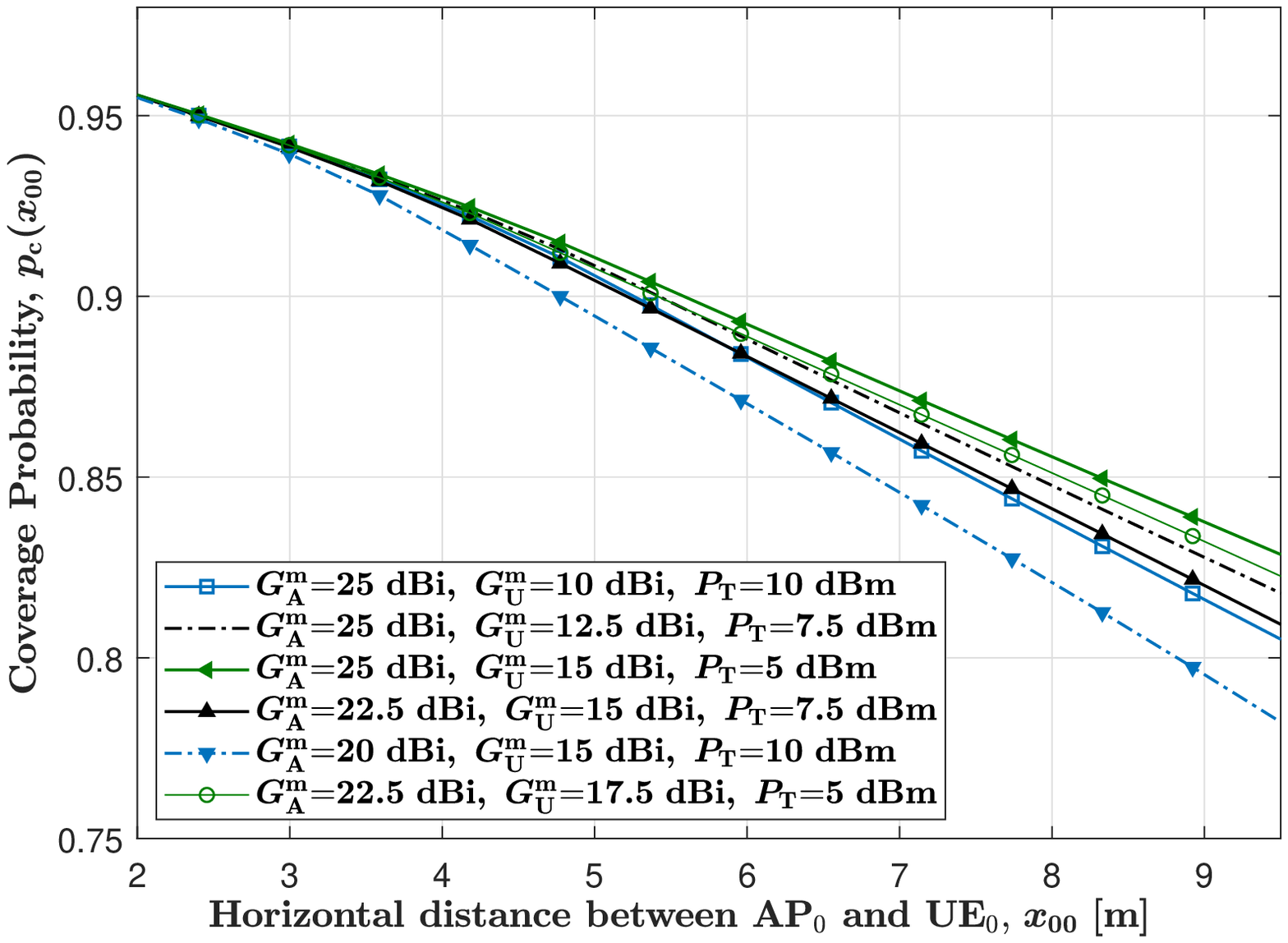}   \vspace{-11mm}
    \caption{The coverage probability versus the UE$_{0}$-AP$_{0}$ link distance  for the typical indoor environment with different antenna gains at UEs and APs.}
    \label{ResFig:Cove}
  \end{minipage}
  \hfill
    \begin{minipage}[h]{0.485\columnwidth}
   \vspace{-3mm}
   \includegraphics[width=\columnwidth]{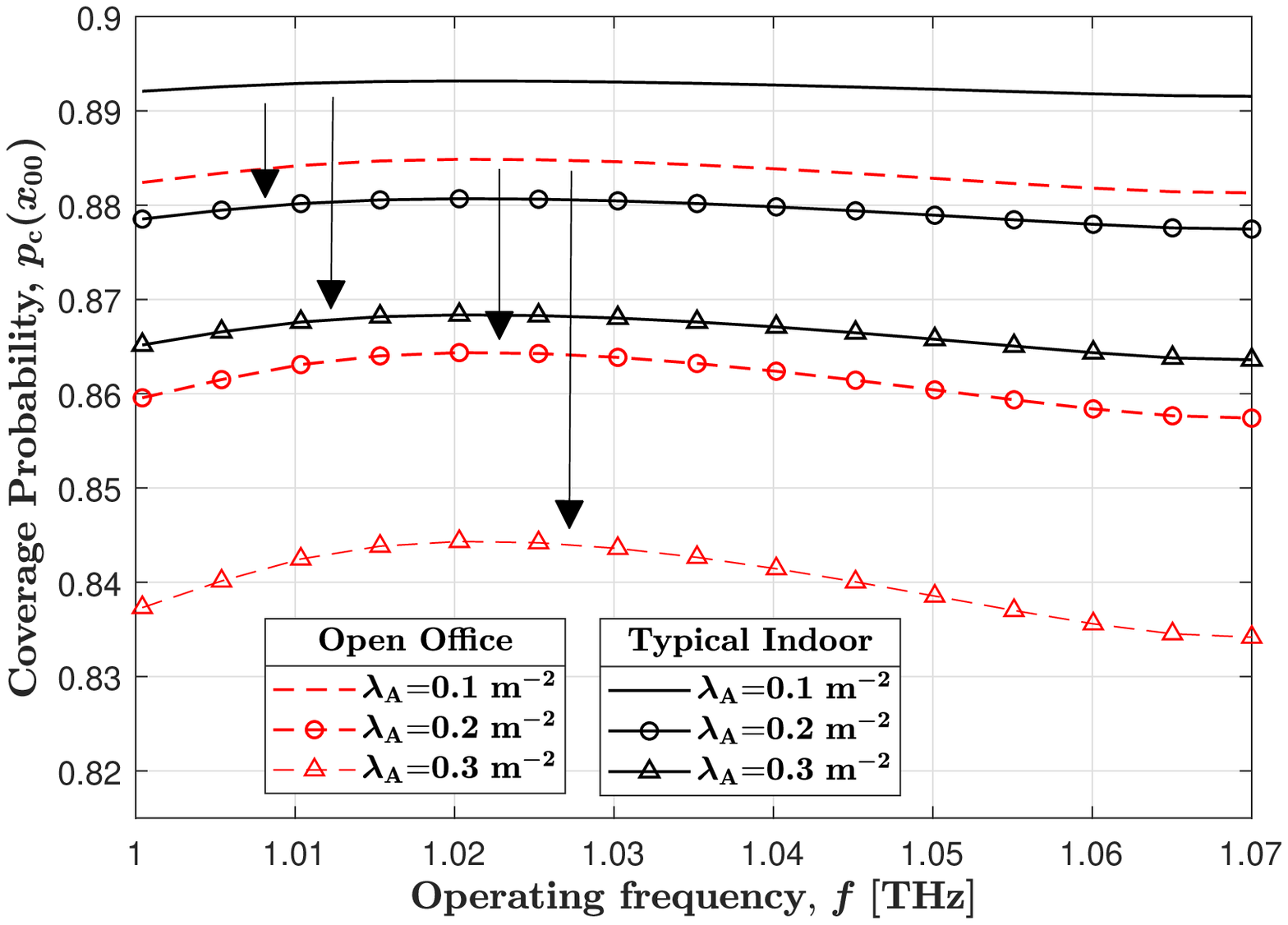} \vspace{-11mm}
    \caption{The coverage probability versus the operating frequency for different values of AP densities.}
    \label{ResFig:Covc}
  \end{minipage}
      \vspace{-5mm}
\end{figure}

Fig.~\ref{ResFig:Cove} plots the coverage probability versus $x_{00}$ for different antenna main lobe gains at APs and UEs, i.e., $G_{\textrm{A}}^{\textrm{m}}$ and $G_{\textrm{U}}^{\textrm{m}}$.
In this figure, we keep $P_{\textrm{T}}G_{\textrm{A}}^{\textrm{m}}G_{\textrm{U}}^{\textrm{m}}$ unchanged for the sake of a fair comparison.
We first observe that the improvement in the coverage probability when  $G_{\textrm{A}}^{\textrm{m}}$ and $G_{\textrm{U}}^{\textrm{m}}$ are increased is marginal for low $x_{00}$ , but is  noticeably high for larger $x_{00}$.
This is due to the fact that the deterioration in the coverage probability caused by interference and the blockage is marginal for small $x_{0}$, but increases when $x_{0}$ becomes large in Fig. \ref{ResFig:Cova}. Thus, the opportunity for coverage improvement by increasing the antenna gains is higher for larger $x_{00}$.
This reveals that the coverage performance of THzCom systems can be improved by increasing the antenna directivity at both the APs and the UEs for larger $x_{00}$.
Second, observing the curves with the same $P_{\textrm{T}}$, we find that the coverage probability improvement brought by increasing $G_{\textrm{A}}^{\textrm{m}}$ is higher than that brought by increasing $G_{\textrm{U}}^{\textrm{m}}$. \emph{This implies that it is more worthwhile to increase the antenna directivity at the APs than at the UEs, to produce a more reliable THzCom system.}

\subsubsection{\underline{Impact of Operating Frequency and AP Densities}}

Fig. \ref{ResFig:Covc} plots the coverage probability versus the operating frequency for different values of AP densities in the first transmission window above $1~\textrm{THz}$, when $x_{00}=6~\textrm{m}$. First, we observe that the coverage probability remains almost unchanged throughout the transmission window, despite  the path loss   varying  drastically within the transmission window as shown in \cite[Fig. 3]{akramICC2020} due to frequency-dependent absorption loss.
This is due to the fact when the operating frequency is changed, while the received power from the desired link changes, the interference power also changes in a similar manner, thereby leading to coverage to remain almost unchanged throughout the transmission window.
Second, we observe that the coverage probability significantly decreases when the density of APs becomes higher for both typical indoor and open office, due to the increased impact from interferers.
Third, we observe that the deterioration in the coverage probability  due to the increased density of APs is less for a typical indoor environment than that of an open office environment.
This is due to the fact that when there are wall blockers in a typical indoor environment, the likelihood of interference signals being blocked becomes higher, which leads to better coverage probability.
Second and third observations above demonstrate that the network densification deteriorates the reliability of THzCom systems, and that it is necessary to carefully select the AP densities to obtain the desired reliability performance depending on the communication environment of interest.


\vspace{-5mm}
\section{Conclusions}\label{sec:conclusions}
\vspace{-1mm}

In this work, we formulated a tractable analytical framework to evaluate the coverage performance of a typical user in an indoor  THzCom environment. We first modeled a realistic 3D THz system, where we considered the unique molecular absorption loss at the THz band, 3D directional antennas at both UEs and APs, and the interference from the nearby APs. Differing from the existing THz studies, we considered the joint impact of the blockage caused by the user itself, moving humans and wall blockers, as well as the effect of the vertical heights of the THz devices.
We then derived the blockage and hitting probabilities that form the basis of the coverage analysis. Thereafter, we derived a new expression for the coverage probability using the dominant interferer analysis for both the typical indoor and the open office environment.
Using numerical results, we validated our analysis and demonstrated that the hitting probability and the coverage probability are significantly overestimated and underestimated, respectively, when the impact of vertical heights of communication entities are ignored in the analysis.
We also found that increasing the antenna directivity at APs brings a larger coverage improvement than increasing the antenna directivity at UEs.


\appendices

\vspace{-3mm}
\section{Proof of Lemma \ref{Lem:pLS}}\label{app:Derive_pLS}
\vspace{-1mm}

The average of the number of walls that intersect the link between AP$_{i}$ and UE$_{j}$ for the binary wall orientation is given by \vspace{-2mm}
\begin{equation}\label{Equ:pLSproof1}
 \varpi_{ij}^{\textrm{W}}=\lambda_{\textrm{W}}\mathbb{E}\left[L_{\textrm{W}}\right] \xi(\phi_{ij})x_{ij} ,
\end{equation}
where $\xi(\phi_{ij})=\frac{1}{2}\left(\vert\sin(\phi_{ij})\vert+\vert\cos(\phi_{ij})\vert\right) $ with $\phi_{ij}$ being the angle that the projection of the UE$_{j}$-AP$_{i}$ link  onto the horizontal plane forms with a reference line in the horizontal plane\cite{IPN1_Wall}. Hence, considering the void probability of walls existing within the link between AP$_{i}$ and UE$_{j}$, we obtain $p_{\textrm{LoS,W}}(x_{ij})$ as \vspace{-3mm}
\begin{align}\label{Equ:pLSproof2}
   p_{\textrm{LoS,W}}(x_{ij})&=e^{-\varpi_{ij}^{\textrm{W}} }= e^{- \lambda_{\textrm{W}}\mathbb{E}\left[L_{\textrm{W}}\right] \frac{1}{2} \left(\vert\sin(\phi_{ij})\vert+\vert\cos(\phi_{ij})\vert\right)x_{ij} }.
\end{align} 
We note that the analysis for the coverage probability is mathematically intractable if the expression for $p_{\textrm{LoS,W}}(x_{ij})$ in \eqref{Equ:pLSproof2} is in its current form.
To address this intractability, we use the fact that $\phi_{ij}$ is uniformly distributed between $-\pi$ and $\pi$, and approximate $p_{\textrm{LoS,W}}(x_{ij})$ for the binary wall orientation as $p_{\textrm{LoS,W}}(x_{ij})   = e^{- \lambda_{\textrm{W}}\mathbb{E}\left[L_{\textrm{W}}\right] \mathbb{E}_{\phi_{ij}}\left[\xi(\phi_{ij})\right] x_{ij}  }$,
where $\mathbb{E}_{\phi_{ij}}\left[\xi(\phi_{ij})\right]=\int_{-\pi}^{\pi}\xi(\phi_{ij}) \frac{1}{2\pi}d\phi_{ij}=\frac{2}{\pi}$. This approximation allows us to obtain  \eqref{Equ:pLS}.

\vspace{-3mm}
\section{Proof of Lemma \ref{Lem:PDFv}}\label{app:Derive_PDFv}
\vspace{-1mm}

\enlargethispage{0.4 cm}


Recall that AP$_{i}$ is assumed to select its associated UE, UE$_{i}$, such that the link between the AP$_{i}$ and UE$_{i}$ is not blocked by wall blockers. Thus, to formulate $f_{x}(x_{ii})$, we first find the PDF of the distance to a random UE located around AP$_{i}$,  denoted by $f_{\tilde{x}}(\tilde{x}_{ii})$, as
\begin{equation}\label{Equ:PDFvProof1}
f_{\tilde{x}}(\tilde{x}_{ii})=\begin{cases}
 \frac{2 \tilde{x}_{ii}}{R_{\textrm{T}}} , &0\leq \tilde{x}_{ii}\leq R_{\textrm{T}},\\
0, &\textrm{otherwise}.
\end{cases}
\end{equation}
Next, using \eqref{Equ:PDFvProof1} and Lemma \ref{Lem:pLS}, we express $f_{x}(x_{ii})$ as
\begin{equation}\label{Equ:PDFvProof2}
  f_{x}(x_{ii})= \tilde{\varrho}f_{\tilde{x}}(\tilde{x}_{ii}) e^{-\eta_{\textrm{W}} x_{ii}},
\end{equation}
where $\tilde{\varrho}$ is the unknown constant that is utilized to ensure $\int_{0}^{\infty}f_{x}(x_{ii}) dx_{ii}=1$. Thereafter, substituting \eqref{Equ:PDFvProof1} in \eqref{Equ:PDFvProof2}, and applying \cite[Eq (2.322.1)]{IntegralBook}
, we obtain \eqref{Equ:PDFv}.

To find the expression for $R_{\textrm{T}}$ in \eqref{Equ:RT}, we ensure that the signal-to-noise ratio (SNR) at UE$_{i}$ when the link between AP$_{i}$ and UE$_{i}$ is not blocked is above $\tau$ for all possible values of $x_{ii}$ \cite{akramICCWS2020}.
Therefore, we let the SNR when the UE-AP distance is $R_{\textrm{T}}$ equal $\tau$ to obtain \eqref{Equ:RT}.

\section{Proof of Proposition \ref{Propos:pHV}}\label{app:Derive_PHV}

In this appendix, without loss of generality, we prove Proposition \ref{Propos:pHV} for $j=0$.
Let denote $\psi_{ij}$ as the angle that the link between AP$_{i}$ and UE$_{j}$ forms with the horizontal plane
as shown in Fig.~\ref{Fig:InfModel}.
For UE$_{0}$  to be within the vertical beamwidth of AP$_{i}$, $\psi_{ii}$ needs to satisfy $\psi_{i0}-\frac{\varphi_{\textrm{A,V}}}{2}\leq\psi_{ii} \leq\psi_{i0}+\frac{\varphi_{\textrm{A,V}}}{2}$.
Therefore, $p_{\textrm{hp,V}}(x_{i0})$ is obtained as 
\begin{equation}\label{Equ:pHVProof1}
p_{\textrm{\textrm{H,V}}}(x_{ii})=\int_{\psi_{i0}-\frac{\varphi_{\textrm{A,V}}}{2}}
^{\psi_{i0}-\frac{\varphi_{\textrm{A,V}}}{2}}f_{\psi}(\psi_{ii})d\psi_{ii},
\end{equation}
where $f_{\psi}(\psi_{ii})$ is the PDF of $\psi_{ii}$.

To formulate $f_{\psi}(\psi_{ii})$, we use the transformation $x_{ii} =\hbar\cot(\psi_{ii})$ into Lemma \ref{Lem:PDFv} to arrive at
\begin{equation}\label{Equ:PDFbeta}
f_{\psi}(\psi_{ii})=\begin{cases}
\varrho \hbar^2\cot(\psi_{ii})\csc^2(\psi_{ii})e^{-\textrm{H}\eta_{\textrm{W}}\cot(\psi_{ii})}, & \bar{\psi} \leq \psi_{ii}\leq\frac{\pi}{2},\\
0, &\textrm{otherwise}.
\end{cases}
\end{equation}
Finally, by substituting \eqref{Equ:PDFbeta} into \eqref{Equ:pHVProof1} and solving the resultant integral by jointly applying  \cite[Eq (2.02.5)]{IntegralBook} with \cite[Eq (2.521.1)]{IntegralBook}, we obtain \eqref{Equ:pHV} for $j=0$.

\section{Proofs of Propositions 3 and 4}\label{app:Derive_Prop234}

\subsection{Proof of Proposition 3}\label{app:Derive_Prop3}

We note that the \emph{near dominant interferers} lead to outage at UE$_{0}$ when their main lobe or the side lobes are facing UE$_{0}$.
Despite that the locations of all the interferers follow a homogeneous PPP with the density $\lambda_{\textrm{A}}$, the interferers that are in LoS with UE$_{0}$ within the region $\mathcal{A}^{\textrm{N}}$ follow a non-homogeneous PPP, due to the distant-dependent nature of the blockage probability.
Thus, using Corollary \ref{Corol:pL}, we probabilistically thin the original homogeneous PPP  to obtain  $\Lambda_{\Phi^{\textrm{N}}}$ as
\begin{align}\label{Equ:densityNProof}
\Lambda_{\Phi^{\textrm{N}}}&=\iint_{ \mathcal{A}^{\textrm{N}}} \lambda_{\textrm{A}}p_{\textrm{LoS}}(x) x d\textrm{A}  \notag \\
&=\int_{0}^{v_{\textrm{s},\textrm{s},1}}\int_{0}^{\varphi_{\textrm{U,H}}}\lambda_{\textrm{A}}\zeta e^{-\eta x} x d\theta  dx + \int_{\check{x}_{00}}^{v_{\textrm{m},\textrm{s}}}\int_{0}^{\varphi_{\textrm{U,H}}}\lambda_{\textrm{A}}\zeta e^{-\eta x} x d\theta  dx \notag\\
& ~~~~~~~~~~ +\int_{\hat{x}_{00}}^{v_{\textrm{s},\textrm{s},2}}\int_{0}^{\varphi_{\textrm{U,H}}}\lambda_{\textrm{A}}\zeta e^{-\eta x} x d\theta  dx + \int_{0}^{D_{\textrm{s},\textrm{s}}}\int_{\varphi_{\textrm{U,H}}}^{2\pi-\omega}\lambda_{\textrm{A}}\zeta e^{-\eta x} x d\theta  dx.
\end{align}
Thereafter, we apply \cite[Eq (2.322.1)]{IntegralBook} into \eqref{Equ:densityNProof} to obtain \eqref{Equ:densityN}.

\subsection{Proof of Proposition 4}\label{app:Derive_Prop4}

Unlike the \emph{near dominant interferers}, for the \emph{far dominant interferers} to cause outage, their main lobe should be facing UE$_{0}$.
Thus, using Proposition \ref{Propos:pHV} and Corollary \ref{Corol:pL},
 we probabilistically thin the original homogeneous PPP  to obtain  $\Lambda_{\Phi^{\textrm{F}}}$ as
 \begin{align}\label{Equ:densityFProof}
\Lambda_{\Phi^{\textrm{F}}}&=\iint_{ \mathcal{A}^{\textrm{F}}} \lambda_{\textrm{A}} p_{\textrm{hp}}(x) p_{\textrm{LoS}}(x) x d\textrm{A}   \notag \\
&=\frac{\lambda_{\textrm{A}} \zeta \varphi_{\textrm{A,H}} \varphi_{\textrm{U,H}}}{2\pi} \Big[\int_{v_{\textrm{m},\textrm{s}}}^{v_{\textrm{m},\textrm{m}}} \!\!p_{\textrm{hp,V}}(x) e^{-\eta x} x  dx{+} \int_{v_{\textrm{s},\textrm{s},1}}^{v_{\textrm{s},\textrm{m},1}} \!\!p_{\textrm{hp,V}}(x) e^{-\eta x} x  dx {+}\int_{v_{\textrm{s},\textrm{s},2}}^{v_{\textrm{s},\textrm{m},2}} \!\!p_{\textrm{hp,V}}(x) e^{-\eta x} x  dx  \Big]  \notag \\
& ~~~~~~~  + \frac{\lambda_{\textrm{A}} \zeta \varphi_{\textrm{A,H}} \left(2\pi -  \varphi_{\textrm{U,H}}-\omega\right)}{2\pi} \int_{D_{\textrm{s},\textrm{s}}}^{D_{\textrm{s},\textrm{m}}} p_{\textrm{hp,V}}(x) e^{-\eta x} x  dx,
\end{align}
which leads to \eqref{Equ:densityF}.

\section{Derivation of $\digamma(a,b)$}\label{app:F(a,b)}

Let us denote $I_{1}^\pm(x)$ and $I_{2}(x)$ as the integrals given by \begin{equation} \label{Equ:DerI1pm}
    I_{1}^{\pm}(x)= \int \frac{\hbar^{2}}{R_{\textrm{T}}^{2}}\cot^2\left(\frac{\varphi_{\textrm{A,V}}}{2}\pm \arctan\left(\frac{\hbar}{x}\right)\right) e^{-\eta_{\textrm{B}} x} x  \;dx
\end{equation}
and
\begin{equation} \label{Equ:DerI2}
    I_{2}(x)= \int e^{-\eta_{\textrm{B}} x} x \; dx,
\end{equation}
respectively. Therefore, $\digamma_{\!\!\textrm{o}}(a,b)$ is obtained as in \eqref{Equ:F(a,b)} by performing the integral $\digamma_{\!\!\textrm{o}}(a,b)=\int_{a}^{b} p_{\textrm{hp,V}}(x) e^{-\eta_{\textrm{B}}
 x} x  \;dx$ using  \eqref{Equ:DerI1pm}, \eqref{Equ:DerI2}, and Corollary \ref{Corol:pHVOO}.

We next present the proof of $I^{+}(x)$ when $\eta_{\textrm{B}} = 0$.
To this end, we expand $I^{+}(x)$ when $\eta_{\textrm{B}} = 0$, $I_{1}^{+}(x)\vert_{\eta_{\textrm{B}}=0}$,  using \cite[Eq (1.313.9)]{IntegralBook} as
\begin{align} \label{Equ:DerI1p2}
  I_{1}^{+}(x)\vert_{\eta_{\textrm{B}}=0}   & =\frac{\hbar^{2}}{R_{\textrm{T}}^{2}} \int \frac{\cos^2(\frac{\varphi_{\textrm{A,V}}}{2})x^3-2 \hbar\sin(\frac{\varphi_{\textrm{A,V}}}{2})\cos(\frac{\varphi_{\textrm{A,V}}}{2})x^2+\hbar^2\sin^2(\frac{\varphi_{\textrm{A,V}} }{2})x}{\left( \hbar \cos( \frac{\varphi_{\textrm{A,V}}}{2}) + x\sin(\frac{\varphi_{\textrm{A,V}}}{2})\right)^2}   dx.
\end{align}
Thereafter, we apply \cite[Eq (2.111.4)]{IntegralBook}
for $z=\vartheta+ \varsigma x$ and $n=1,2,3$, into \eqref{Equ:DerI1p2} and rearrange the resulting terms to arrive at \eqref{Equ:I1pmab} when $\eta_{\textrm{B}}=0$.

We next present the proof of $I^{+}(x)$ when $\eta_{\textrm{B}}\neq 0$. To this end, we first apply the integration by parts formula \cite[Eq (2.02.5)]{IntegralBook}  into \eqref{Equ:DerI1pm}, which leads to
\begin{align}
 \label{Equ:Ipmab5}
    I_{1}^{+}(x)\vert_{\eta_{\textrm{B}}\neq0}&=\!\! \int  \!\! e^{-\eta_{\textrm{B}} x} \times\frac{d}{dx}\left\{I_{1}^+(x)\vert_{\eta_{\textrm{B}}=0}\right\} dx  {=} \underbrace{I_{1}^+(x)\vert_{\eta_{\textrm{B}}=0} e^{-\eta_{\textrm{B}} x}}_{\substack{T_{1}}} + \!\!\underbrace{\int \!\! \eta_{\textrm{B}} I_{1}^+(x)\vert_{\eta_{\textrm{B}}=0} e^{-\eta_{\textrm{B}} x}   dx }_{\substack{T_{2}}}.
\end{align}
Therefore, to obtain $T_{2}$ in \eqref{Equ:Ipmab5}, we first apply \cite[Eq (2.02.5)]{IntegralBook} into \cite[Eq (3.353.1)]{IntegralBook}, which leads to \vspace{-3mm}
\begin{equation}\label{Equ:DerI1pm12}
\int \ln(z)e^{-kx} dx=\frac{1}{k} \left(e^\frac{k \vartheta}{\varsigma}\textrm{Ei}\left[-\frac{k z}{\varsigma} \right]-e^{k x}\ln(z)\right),
\end{equation}
where $z=\vartheta+\varsigma x$ and $k\neq 0$. Thereafter, applying \eqref{Equ:DerI1pm12}, \cite[Eq (2.321.2)]{IntegralBook}, and \cite[Eq (2.02.5)]{IntegralBook} to $T_{2}$, we obtain
\begin{align}
 \label{Equ:Ipmab6}
 &T_{2}\!\! = \frac{ \eta_{\textrm{B}} \hbar^{2}}{R_{\textrm{T}}^{2}}\Bigg[ -\frac{\cot^2(\frac{\varphi_{\textrm{A,V}}}{2})}{2\eta_{\textrm{B}}^3} e^{-\eta_{\textrm{B}} x} \left(2+2\eta_{\textrm{B}} x+\eta_{\textrm{B}}^2 x^2\right)+ \frac{\hbar \csc^4(\frac{\varphi_{\textrm{A,V}}}{2})\sin(\varphi_{\textrm{A,V}})}{\eta_{\textrm{B}}^2} e^{-\eta_{\textrm{B}} x} (1+\eta_{\textrm{B}} x)  \notag \\
& {+} \hbar^3 \csc^6(\frac{\varphi_{\textrm{A,V}}}{2}) \cos(\frac{\varphi_{\textrm{A,V}}}{2})  e^{\eta_{\textrm{B}} \hbar \cot(\frac{\varphi_{\textrm{A,V}}}{2})}  \textrm{Ei}\left[{-} \eta_{\textrm{B}} x {-} \eta_{\textrm{B}} \hbar \cot(\frac{\varphi_{\textrm{A,V}}}{2}) \right]{+}\frac{\hbar^2}{\eta_{\textrm{B}}} \left(2{+}\cos(\varphi_{\textrm{A,V}}) \right) \csc^4(\frac{\varphi_{\textrm{A,V}}}{2}) \notag \\
    & ~~~~~\times \Big(e^{\eta_{\textrm{B}} \hbar \cot(\frac{\varphi_{\textrm{A,V}}}{2})}  \textrm{Ei}\left[{-} \eta_{\textrm{B}} x {-} \eta_{\textrm{B}} \hbar \cot(\frac{\varphi_{\textrm{A,V}}}{2}) \right]{-}\ln(\hbar \cos(\frac{\varphi_{\textrm{A,V}}}{2}
){+} x \sin(\frac{\varphi_{\textrm{A,V}}}{2}
))e^{{-}\eta_{\textrm{B}} x} \Big)  \Bigg].\!\!\!\!
\end{align}
Finally, by substituting \eqref{Equ:Ipmab6} into \eqref{Equ:Ipmab5} and expanding $T_{1}$, we arrive at \eqref{Equ:I1pmab} when $\eta_{\textrm{B}} \neq0$.

The proof of $I_{1}^-(x)$ is omitted since it is similar to that of $I_{1}^+(x)$. In addition, we clarify that $I_{2}(x)$ in \eqref{Equ:I2ab} is obtained by performing the integral in \eqref{Equ:DerI2} by applying \cite[Eq (2.322.1)]{IntegralBook}.

\bibliographystyle{IEEEtran}
\bibliography{JSAC2021_ArXivV2}

\end{document}